\documentclass[prd,aps,preprintnumbers,showpacs,amsfonts,floatfix,nofootinbib]{revtex4}
\usepackage{epsfig,latexsym,graphicx,graphics}

\begin{document}
\begin{titlepage}

\preprint{Napoli Preprint DSF-2003-38}

\title{
$J/\psi$ dissociation cross sections in a relativistic quark model}

\author{\textbf{Mikhail A. Ivanov}}
\affiliation{Bogoliubov Laboratory of Theoretical Physics, \\
Joint Institute for Nuclear Research, 141980 Dubna, Russia}

\author{\textbf{Juergen G. K\"{o}rner}}
\affiliation{Institut f\"{u}r Physik, \\
Johannes Gutenberg-Universit\"{a}t, \\
D-55099 Mainz, Germany}

\author{\textbf{Pietro Santorelli}}
\affiliation{Dipartimento di Scienze Fisiche, Universit{\`a} di
Napoli "Federico II", Italy\\
and Istituto Nazionale di Fisica Nucleare, Sezione di Napoli, Italy}

\begin{abstract}
We calculate the amplitudes and the cross sections of the charm
dissociation processes $ J/\psi + \pi \to   D \overline D,
D^\ast\overline D \,( \overline {D^\ast} D), D^\ast\overline
{D^\ast}$ within a relativistic constituent quark model. We
consistently account for the contributions coming from both the
box and triangle diagrams that contribute to the dissociation
processes. The cross section is dominated by the $D^\ast\overline
D$ and $D^\ast\overline {D^\ast}$ channels. When summing up the
four channels we find a maximum total cross section of about 2.3
mb at $\sqrt{s}\approx 4.1$ GeV. We compare our results to the
results of other model calculations.
\end{abstract}

\maketitle
\end{titlepage}

\section{Introduction}
\label{s:introduction}

The analysis of the $J/\psi$ dissociation cross section is
important for understanding the suppression of $J/\psi$
production observed in Pb-Pb collisions by the NA50 Collaboration
at the CERN Super Proton Synchrotron (SPS) \cite{NA50}. There are a number of theoretical
calculations on the $c\bar c$ + light hadron cross sections (see,
e.g., the review Ref.~\cite{Barnes:2003vt}). However, they give
widely divergent results, which implies that one is still far away
from a real understanding of the scattering mechanism. The
nonrelativistic quark model has been applied in
\cite{Martins:1994hd} and
\cite{Wong:1999zb,Wong:2001td,Barnes:2003dg} for the calculation
of the cross sections for the dissociation processes $c\bar
c+q\bar q\to c\bar q + q\bar c$. The calculated
cross sections for the reactions $J/\psi + \pi \to D\overline D $,
$D^\ast\overline D$, $D \overline {D^\ast}$, $D^\ast\overline
{D^\ast}$ have the following common features: they rise very fast
from zero at threshold  to a maximum value and finally fall off
due the Gaussian form of the potential. The magnitude of the
maximum total cross section was found to be $\approx 7$ mb at
$\sqrt{s}\approx 4.1$ GeV in \cite{Martins:1994hd} and a somewhat
smaller value of $\approx 1.4$ mb at $\sqrt{s}\approx 3.9$ GeV in
\cite{Wong:1999zb,Wong:2001td,Barnes:2003dg}.

Another approach to studying the charm dissociation process started
with the model proposed by Matinian and M\"{u}ller \cite{Matinian:1998cb}.
They assumed that the dissociation cross section is dominated, in the
$t$ channel, by the D meson exchange. A generalization of this approach
can be found in \cite{Haglin:1999xs,Haglin:2003fh,Lin:1999ad,
Oh:2000qr,Oh:2002vg,Ivanov:2001th} where an effective chiral SU(4)
Lagrangian was employed. Such an approach seems to us quite
problematic for the following reasons: (i) SU(4) is a badly broken
symmetry, and (ii) some of the couplings in the chiral SU(4)
Lagrangian are unknown. Nevertheless, this is a relativistic
approach that allows one to study  the above processes in a
systematic fashion. In this framework, the dissociation cross
section of $J/\psi$ by light hadrons is predicted to be, near the
physical threshold, in the range 1-10 mb. Moreover, it is
interesting to refer to the paper \cite{Oh:2002vg}, where
the $J/\psi$ dissociation by $\pi$ and $\rho$ mesons was examined
in the meson exchange model \cite{Oh:2002vg} and compared to the
quark interchange model. The authors of this paper found that the meson
exchange model
could give predictions similar to those of the quark interchange models,
not only for the magnitudes but also for the energy dependence of
the low-energy dissociation cross sections of the $J/\psi$ by
$\pi$ and $\rho$ mesons.

It appears that the microscopic quark nature of hadrons is
important in the charm dissociation processes. The first step is
to calculate the relevant form factors corresponding to the triple
and quartic meson vertices in the kinematical region of the
dissociation reaction. QCD sum rules have been used in Refs.~
\cite{Navarra:2001jy,Navarra:2001ju,Matheus:2002nq,Duraes:2002ux}
to evaluate those form factors and to determine the charm cross
section. The cross section was found to be about 1 mb at
$\sqrt{s}\approx 4.1$ GeV with a monotonic growth when the energy is
increased~\cite{Duraes:2002ux}.

An approach based on the 1/N expansion in QCD combined with the Regge theory
gives, for the total dissociation cross section, a value of a
few millibarns near to $\sqrt{s}\approx 4$ GeV \cite{1/N_Regge}.

We also mention the work of Deandrea {\it et al.}, where the strong
couplings $J/\psi D^\ast D^\ast$ and $J/\psi D^\ast D^\ast \pi$
were evaluated in the constituent quark model
\cite{Deandrea:2003pv}. Finally, an extension of
the finite-temperature Dyson-Schwinger equation approach to
heavy mesons and its application to the reaction $J/\psi +\pi \to
D + \overline D$ was considered in \cite{Blaschke:2000zm}.

We employ a relativistic quark model \cite{model-1} to  calculate
the charm dissociation amplitudes and cross sections. This model
is based on an effective Lagrangian which describes the coupling
of hadrons $H$ to their constituent quarks. The coupling strength
is determined by the compositeness condition $Z_H=0$ \cite{z=0}
where $Z_H$ is the wave function renormalization constant of the
hadron $H$.
One starts with an effective Lagrangian written down in terms of
quark and hadron fields. Then, by using  Feynman rules, the
S-matrix elements describing the hadronic interactions are given
in terms of a set of quark diagrams. In particular, the
compositeness condition enables one to avoid a double counting of
the hadronic degrees of freedom. The approach is self-consistent
and universally applicable. All calculations of physical
observables are straightforward. The model has only a small set of
adjustable  parameters given by  the values of the constituent
quark masses and the scale parameters that define the size of the
distribution of the constituent quarks inside a given hadron. The
values of all fit parameters are within the window of
expectations.

The shape of the  vertex functions and the  quark  propagators can
in principle  be found from an analysis of the Bethe-Salpeter and
Dyson-Schwinger equations  as was done, e.g., in
\cite{Ivanov:1998ms}. In this paper, however, we choose a
phenomenological approach where the vertex functions are modeled
by a Gaussian form, the size parameter of which is determined by a
fit to the leptonic and radiative decays of the lowest-lying
light, charm, and bottom mesons. For the quark propagators we use a
local representation.

We calculate the amplitudes and the cross sections of the charm
dissociation processes
\begin{eqnarray*}
J/\psi + \pi & \to & D + \overline D, \\
J/\psi + \pi & \to & D^\ast+ \overline D \,\,\,( \overline {D^\ast}+D), \\
J/\psi + \pi & \to & D^\ast + \overline {D^\ast}.
\end{eqnarray*}
These processes are described by both box and resonance diagrams
which can be calculated straightforwardly in our approach. We
compare our results with the results of other studies.

The layout of the paper is as follows. In Sec. II we briefly
discuss our relativistic quark model. In Sec. III we outline the
calculational technique of the arbitrary $n$-point one-loop
diagrams with local propagators and Gaussian vertex functions. We
give explicit result for the triangle and box diagrams which are
the building blocks of the charm dissociation amplitudes. In Sec.
IV we calculate the charm dissociation amplitudes and the total
cross sections. In Sec V we perform the numerical analysis and
give our prediction for the cross sections.

\section{The Model}
\label{s:model}

The coupling of a meson $H$ to its constituent quarks $q_1$ and
$\bar q_2 $ is determined by the Lagrangian
\begin{equation}\label{Lagr_str}
{\cal L}_{\rm int}^{\rm Str}(x) = g_H H(x)\int\!\! dx_1
\!\!\int\!\! dx_2 F_H (x,x_1,x_2)\bar q_2(x_2)\Gamma_H\lambda_H
q_1(x_1) \, + {\rm h.c.}\, .
\end{equation}
Here, $\lambda_H$ and $\Gamma_H$ are  Gell-Mann and Dirac matrices
which describe the flavor and spin quantum numbers of the meson
field $H(x)$. The function $F_H$ is related to the scalar part of
the Bethe-Salpeter amplitude and characterizes the finite size of
the meson. To satisfy translational invariance, the function $F_H$
has to satisfy the identity $F_H(x+a,x_1+a,x_2+a)=F_H(x,x_1,x_2)$
for any four-vector $a$. In the following we use a particular form
for the vertex function,
\begin{equation}\label{vertex}
F_H(x,x_1,x_2)=\delta(x - c^1_{12}x_1 - c^2_{12}x_2)
\Phi_H\left((x_1-x_2)^2\right) ,
\end{equation}
where $\Phi_H$ is the correlation function of two constituent
quarks with masses $m_1$, $m_2$ and $c^k_{ij}=m_k/(m_i+m_j)$.

The coupling constant $g_H$ in Eq.~(\ref{Lagr_str}) is determined
by the so-called {\it compositeness condition} originally proposed
in~\cite{z=0} and extensively used in~\cite{model-2}. The
compositeness condition requires that the renormalization constant
of the elementary meson field $H(x)$ is set to zero,
\begin{equation}\label{z=0}
Z_H \, = \, 1 - \, \frac{3g^2_H}{4\pi^2} \,
\tilde\Pi^\prime_H(M^2_H) \, = \, 0,
\end{equation}
where $\tilde\Pi^\prime_H$ is the derivative of the meson mass
operator. In order to clarify the physical meaning of this
condition, we note that  $Z_H^{1/2}$ is also interpreted as the
matrix element between a physical particle state and the
corresponding bare state. For $Z_H=0$ it then follows that the
physical state does not contain the bare one and is described as a
bound state. The interaction Lagrangian in Eq.~(\ref{Lagr_str})
and the corresponding free Lagrangian describe both the
constituents (quarks) and the physical particles (hadrons) which
are  bound states of the quarks. As a result of the interaction,
the physical particle is dressed, i.e., its mass and wave function
have to be renormalized. The condition $Z_H=0$ also effectively
excludes the constituent degrees of freedom from the physical
space and thereby guarantees that a double counting of physical
observables is avoided. The constituent quarks exist in virtual
states only. One of the corollaries of the compositeness condition
is the absence of a direct interaction of the dressed charged
particle with the electromagnetic field. Taking into account both
the tree-level diagram and the diagrams with the self-energy
insertions into the external legs yields a common factor $Z_H$
which is equal to zero. We refer the interested reader to our
previous papers \cite{model-1,model-2,model-3} where these
points are discussed in more detail.

We briefly discuss the introduction of the electromagnetic field
into the nonlocal Lagrangian in Eq.~(\ref{Lagr_str}) in a gauge invariant
manner. This can be accomplished by using the path exponential \cite{Mandelstam}
\begin{eqnarray}
\label{gauging}
{\cal L}_{\rm int}^{\rm Str + em}(x) &=&
g_H H(x)\int\!\! dx_1 \!\!\int\!\!dx_2 F_H (x,x_1,x_2)
\bar q_2(x_2)\, e^{ie_{q_2} I(x_2,x,P)}\,
\Gamma_H\lambda_H\,
e^{-ie_{q_1} I(x_1,x,P)}\,  q_1(x_1),
\end{eqnarray}
where

\begin{equation}
\label{path}
I(x_i,x,P) = \int\limits_x^{x_i} dz_\mu A^\mu(z)
\end{equation}
and where $z^\mu$ is a coordinate point on the path $P$.
At first sight it appears that the results depend on the path $P$
which connects the end points $x$ and $ x_1$ in the path integral
in Eq.~(\ref{path}). However, we need to know only derivatives of such
integrals for the perturbative calculations here.
Therefore, we use a formalism~\cite{Mandelstam}
which is based on the path-independent definition of the derivative of
$I(x,y,P)$:
\begin{eqnarray}\label{path1}
\lim\limits_{dx^\mu \to 0} dx^\mu
\frac{\partial}{\partial x^\mu} I(x,y,P) \, = \,
\lim\limits_{dx^\mu \to 0} [ I(x + dx,y,P^\prime) - I(x,y,P) ]\ ,
\end{eqnarray}
where the path $P^\prime$ is obtained from $P$ by shifting
the end point $x$ by $dx$.
The use of the definition in Eq.~(\ref{path1}) leads to the key rule
\begin{eqnarray}
\label{path2}
\frac{\partial}{\partial x^\mu} I(x,y,P) = A_\mu(x)\ ,
\end{eqnarray}
which in turn states that the derivative of the path integral $I(x,y,P)$ does
not depend on the path P originally used in the definition.
This allows one to construct the perturbation theory in a
consistent way and guarantees the implementation of
charge conservation and the Ward identities.

\begin{figure}[!ht]
\begin{center}
\epsfig{figure=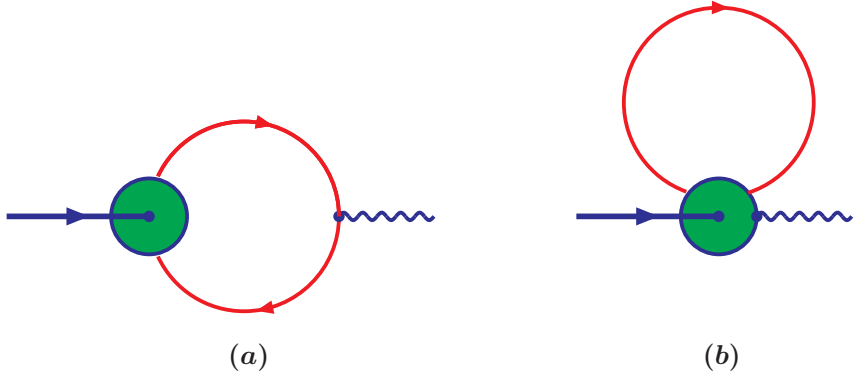,height=5truecm}
\end{center}
\caption{The diagrams corresponding to the $J/\psi\to \gamma$ transition:
(a) the usual one and (b) the "tadpole".} \label{fig:J_gam}
\end{figure}

As an example, we consider the transition $J/\psi\to \gamma$
which is now described by two diagrams in Fig.~\ref{fig:J_gam}:
(a) the standard ``bubble'' and (b) the ``tadpole'' ones.
The total matrix element has a manifestly gauge invariant
form
\begin{equation}
\label{Jpsigam}
M^{\mu\nu}_{J/\psi}(p) =
(g^{\mu\nu}\,p^2-p^\mu p^\nu)\,M_{J/\psi}(p^2)\ ,
\end{equation}
where
\begin{eqnarray*}
M_{J/\psi}(p^2) &=& e_c\frac{3g_{J/\psi}}{4\pi^2}
\frac{1}{p^2}
\int\limits_0^\infty dt\frac{t}{(1+t)^2}\int\limits_0^1 d\alpha
\left\{
y_a\tilde\Phi_{J/\psi}(z_a)-y_b\tilde\Phi_{J/\psi}(z_b)
\right\}\ ,
\\
&&\\
y_a &=&
\left(1+\frac{t}{2}\right)\,m_c^2+
\frac{p^2}{4}\frac{1+3\, t/2-2\,t^3\,\alpha(1-\alpha)}{(1+t)^2}\ ,
\\
z_a &=& t\,[m_c^2-\alpha(1-\alpha)\,p^2]
-\frac{t}{1+t}\left(\frac{1}{2}-\alpha\right)^2\,p^2 \ ,
\\
&&\\
y_b &=&
\left(1+\frac{t}{2}\right)
\left[m_c^2-\frac{\alpha^2}{(1+t)^2}\frac{p^2}{4}\right]\ ,
\\
z_b &=& t\,m^2_c-\left(1-\frac{\alpha}{1+t}\right)\frac{\alpha\,p^2}{4} \ .
\end{eqnarray*}

For the pseudoscalar and vector mesons treated in this paper the
derivatives of the mass operators are written as
\begin{eqnarray}
\tilde\Pi'_P(p^2)&=& \frac{1}{2p^2}\,p^\alpha\frac{d}{d
p^\alpha}\, \int\!\! \frac{d^4k}{4\pi^2i} \tilde\Phi^2_P(-k^2)
{\rm tr} \biggl[\gamma^5 S_1(\not\! k+c^1_{12} \not\!p) \gamma^5
                         S_2(\not\! k-c^2_{12} \not\!p) \biggr],
\nonumber\\
&&\label{Mass-operator}\\
\tilde\Pi'_V(p^2)&=& \frac{1}{3}\biggl[g^{\mu\nu}-\frac{p^\mu
p^\nu}{p^2}\biggr] \frac{1}{2p^2}\,p^\alpha\frac{d}{dp^\alpha}\,
\int\!\! \frac{d^4k}{4\pi^2i} \tilde\Phi^2_V(-k^2) {\rm tr}
\biggl[\gamma^\nu S_1(\not\! k+c^1_{21} \not\!p) \gamma^\mu
                           S_2(\not\! k-c^2_{12}\not\! p)\biggr].
\nonumber
\end{eqnarray}
The leptonic decay constants $f_P$ and $f_V$ are
\begin{eqnarray*}
&& \frac{3g_P}{4\pi^2} \,\int\!\! \frac{d^4k}{4\pi^2i}
\tilde\Phi_P(-k^2) {\rm tr} \biggl[\gamma^\mu \gamma^5 S_1(\not\!
k+c^1_{12} \not\!p) \gamma^5
                       S_2(\not\! k-c^2_{12} \not\!p) \biggr]
=f_P\,p^\mu\, ,
\\
&&\\
&& \frac{3g_V}{4\pi^2} \,\int\!\! \frac{d^4k}{4\pi^2i}
\tilde\Phi_V(-k^2) {\rm tr} \biggl[\gamma^\mu S_1(\not\!
k+c^1_{12} \not\!p) \gamma \cdot\epsilon_V S_2(\not\! k-c^2_{12}
\not\!p) \biggr] =m_V\,f_V\,\epsilon_V^\mu\,.
\end{eqnarray*}
We use free fermion propagators for the valence quarks
\begin{equation}
S_i(\not\! k)=\frac{1}{m_i-\not\! k}
\end{equation}
with an effective constituent quark mass $m_i$. As discussed
in~\cite{model-1,model-2} we assume for the meson mass $M_H$ that
\begin{equation}
\label{conf} M_H < m_{1} + m_{2}
\end{equation}
in order to avoid the appearance of imaginary parts in the
physical amplitudes.  This holds true for the light pseudoscalar
mesons but is no longer true for the light vector mesons. We shall
therefore employ identical  masses for the pseudoscalar mesons and
the vector mesons in our matrix element calculations but use
physical masses in the phase space calculation. This is quite a
reliable approximation for the heavy vector mesons, e.g. $D^\ast$
and $B^\ast$, where the hyperfine splitting between the $D^\ast$
and $D$ and the $B^\ast$ and $B$, respectively, is quite small.

The shape of the  vertex functions and the  quark  propagators can
in principle  be found from an analysis of the Bethe-Salpeter and
Dyson-Schwinger equations  as was done, e.g., in
\cite{Roberts:dr,Ivanov:1998ms}. In this paper, however, we choose
a  phenomenological approach where the vertex functions are
modeled by a Gaussian form, the size parameter of which is
determined by a fit to the leptonic and radiative decays of the
lowest-lying light, charm, and bottom mesons (see
Table~\ref{t:datafit}). Our previous studies of phenomena
involving  the low-lying hadrons have shown that this
approximation is successful and reliable \cite{model-1,model-2}.
We employ a Gaussian for the vertex function of the form
$\tilde\Phi_H(k^2_E) \doteq \exp(- k^2_E/\Lambda^2_H)$, where
$k_E$ is a Euclidean momentum. The size parameters $\Lambda^2_H$
are determined by a fit to experimental data, when available, or
to lattice results for the leptonic decay constants $f_P$ and
$f_V$ where $P=\pi, D, B$ and $V=J/\psi, \Upsilon$. Here we
improve the fit by using the MINUIT code in a least-squares fit.
The values of the fit parameters are displayed  in
Table~\ref{t:fitresults}. The quality of the fit may be assessed
from the entries in Table~\ref{t:datafit}.
\begin{table}[ht]
\def\arraystretch{1.5}
\begin{center}  $
\begin{array}{||c|c|c||c|c|c|}
\hline\hline
    & \mbox{This model} & \mbox{Expt. or lattice} &  &\mbox{This model}& \mbox{Expt. or lattice}\\
\hline\hline
%
f_\pi      & 130.7      & 130.7\pm 0.1\pm 0.36      &
g_{\pi^0\gamma\gamma}   & 0.272 \mbox{  GeV$^{-1}$} & 0.273 \mbox{  GeV$^{-1}$}  \\
\hline
f_K        & 159.8      & 159.8 \pm 1.4\pm 0.44  &
f_{J/\psi} & 405        &  405 \pm 17 \\
\hline
f_D        & 211        &  \begin{array}{c}
                           203 \pm 14\\
                           226 \pm 15 \end{array}  &
f_B        & 182        &\begin{array}{c}
                          173   \pm 23 \\
                          198   \pm 30 \end{array}\\
\hline
f_{D_s}      & 244        & \begin{array}{c}
                             230 \pm  14 \\
                             250 \pm  30\end{array} &
f_{B_s}      & 209        &  \begin{array}{c}
                              200 \pm 20 \\
                              230 \pm 30 \end{array}\\
\hline
f_{B_c}      &  360       &  360   &
f_{\Upsilon} &  710       &  710 \pm 37 \\
\hline

\hline \hline
\end{array}
$    \end{center} \vspace{-0.5truecm} \caption{ \label{t:datafit}
\small The physical quantities used in least-squares fitting
our parameters. The values are taken from the Particle Data Group
\cite{PDB} or from lattice simulations \cite{Ryan}. The value of $f_{B_c}$ is
our average of QCD sum rules calculations \cite{QCDSR}. All
numbers are given in MeV except for $g_{\pi^0\gamma\gamma}$.}
\end{table}
%

\begin{table}[!!h!!]
\def\arraystretch{1.5}
\begin{center}  $
\begin{array}{||c|c||c|c||c|c||}
\hline\hline
\mbox{Quark masses}& \mbox{(GeV)} & \Lambda_H & \mbox{(GeV)} & \Lambda_H & \mbox{(GeV)}\\
\hline\hline
m_u=m_d & 0.223 & \Lambda_\pi   &  1.074 &  \Lambda_{B_c}   & 1.959  \\
\hline
m_s     & 0.356  &  \Lambda_K & 1.514 &  \Lambda_{J/\psi} & 2.622\\
\hline
m_c   & 1.707    & \Lambda_D=\Lambda_{D_s} & 1.844 &  \Lambda_{\Upsilon} & 3.965 \\
\hline
m_b &  5.121   & \Lambda_B=\Lambda_{B_s} & 1.887  & & \\
\hline\hline
\end{array}
$   \end{center} \vspace{-0.5truecm} \caption{\label{t:fitresults}
\small The fitted values of the model parameters.}
\end{table}


\section{Strong triangle and box diagrams}

\begin{figure}[!ht]
\begin{center}
\begin{tabular}{c}
\epsfig{figure=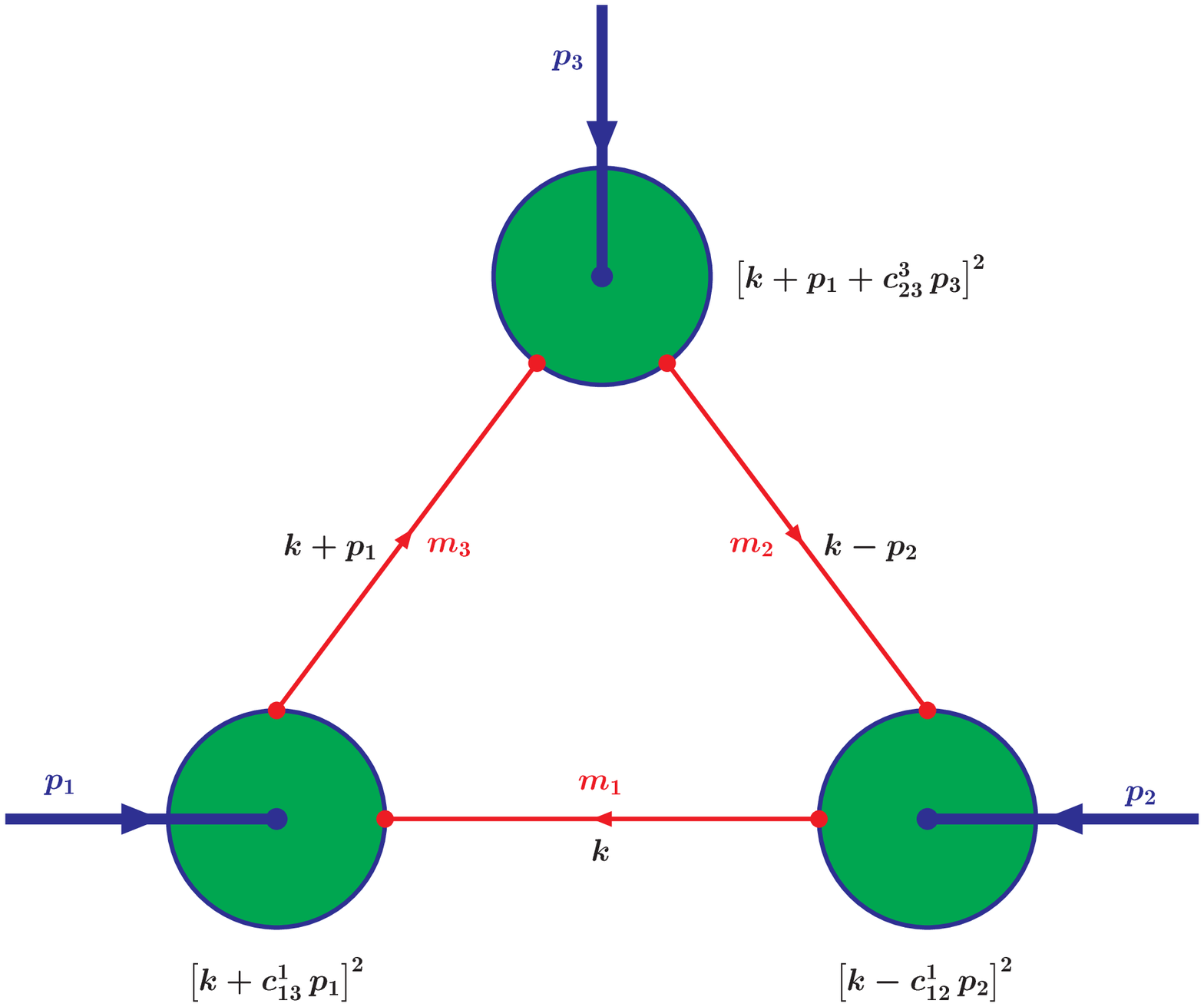,height=9cm} \\
\epsfig{figure=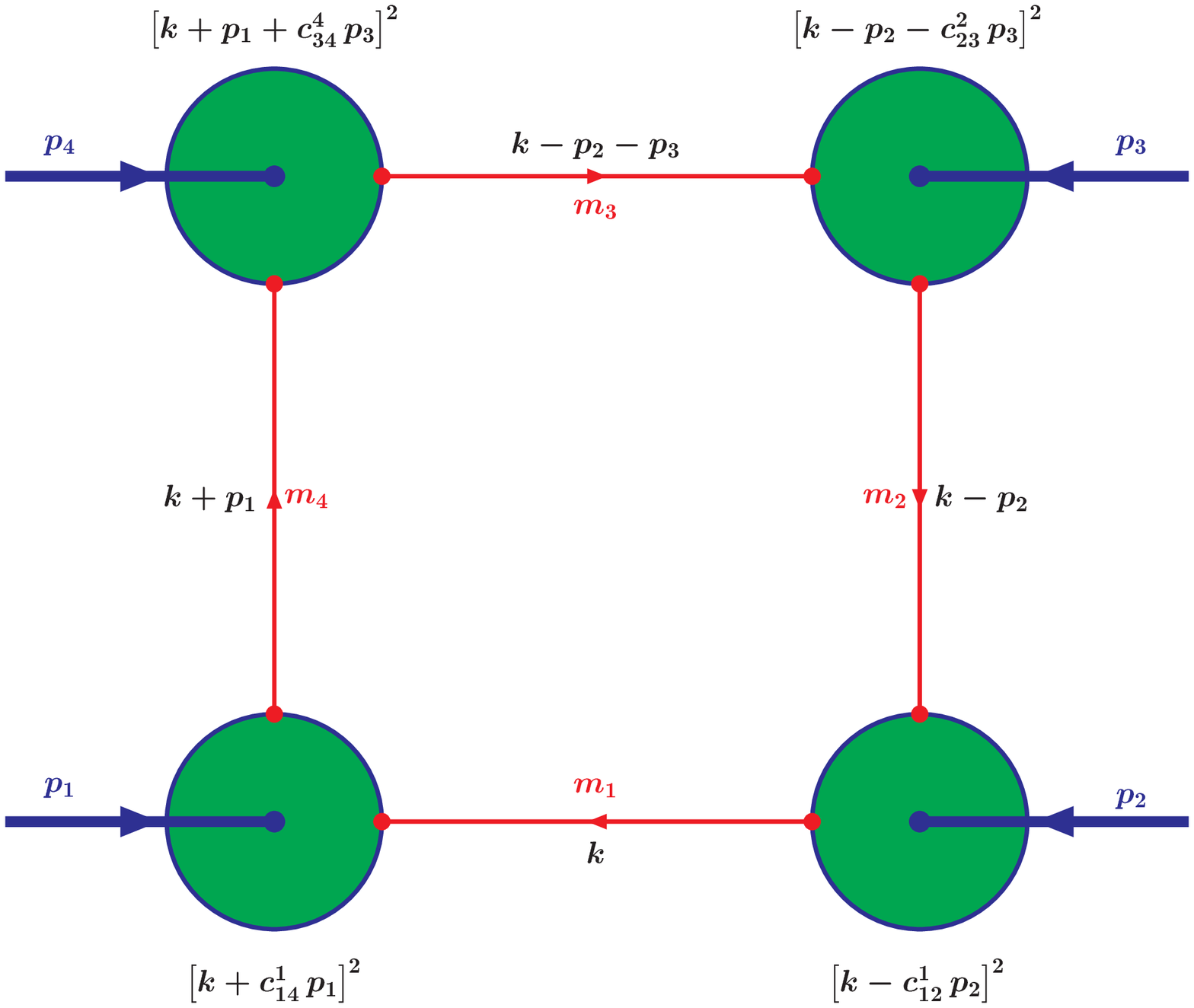,height=9cm} \\
\end{tabular}
\end{center}
\caption{The strong triangle and box diagrams.} \label{fig:loop}
\end{figure}

Transition matrix elements involving composite hadrons are
specified in the model by the appropriate quark diagram. Here, we
give explicit expressions for the integrals corresponding to the
strong triangle and box diagrams shown in Fig.~\ref{fig:loop}.

First, we will make the transformations which are common for all
one-loop diagrams with local propagators and Gaussian vertex
functions. The Feynman integral corresponding to a one-loop
diagram with $n$ propagators and, respectively, $n$ vertex
functions may be written in Minkowskii space as
\begin{equation}
\label{diag_loop} I_n(p_1,...,p_n) = \int\!\! \frac{d^4k}{4\pi^2i}
\, {\rm tr} \left[\prod\limits_{i=1}^n \tilde\Phi_{i} \left( -
(k+v_{i+n})^2\right)\, \Gamma_i\, S_i(\not\! k+\not\! v_i)\right]
\end{equation}
where the vectors $v_i$ are  linear combinations of the external
momenta $p_i$ to be specified later on, $k$ is the loop momentum,
and $\Gamma_i$ are Dirac matrices for the $i$ meson (cf.
Eq.~(\ref{Lagr_str})). The external momenta are all chosen as
ingoing such that one has $\sum\limits_{i=1}^n p_i=0$.

The propagators can be written as
\begin{eqnarray}
\label{prop} S_i(\not\! k+\not\! v_i) &=& (m_i+\not\! k+\not\!
v_i)\cdot\int\limits_0^\infty\! d\beta_i
\,e^{-\beta_i\,(m^2_i-(k+v_i)^2)}\,.
\end{eqnarray}
For the vertex functions one writes
\begin{eqnarray}
\label{vert} \tilde\Phi_{i} \left( - (k+v_{i+n})^2\right) &=&
e^{\beta_{i+n}\cdot(k+v_{i+n})^2} \hspace{1cm} i=1,...,n\, ,
\end{eqnarray}
where the parameters $\beta_{i+n}=s_{i}=1/\Lambda^2_{i}$ are the
size parameters. One can then easily perform the integration over
$k$:
\begin{equation}
 \int\!\! \frac{d^4k}{\pi^2i}\,
\exp\left\{\sum\limits_{i=1}^{2\,n}\beta_i\,(k+v_i)^2\right\} =
\frac{1}{\beta^2} \exp\left\{\frac{1}{\beta} \sum\limits_{1\le i<j
\le 2\,n}\beta_i\,\beta_j\,(v_i-v_j)^2\right\}\, .
\end{equation}
Here, $\beta=\sum\limits_{i=1}^{2\,n}\beta_i$. The numerator
$m_i\, + \!\not\! k\, + \!\not\! v_i$ can be replaced by a
differential operator in the following manner:
\begin{eqnarray}
\label{numer} \int\!\! \frac{d^4k}{\pi^2i}\, \prod\limits_{i=1}^n
\Gamma_i \left(m_i+\not\! k+\not\! v_i\right)e^{\beta k^2+2kr} &=&
\prod\limits_{i=1}^n \Gamma_i \left(m_i+\not\!
v_i+\frac{1}{2}\not\!\partial_r\right)
\cdot\frac{1}{\beta^2}e^{-r^2/\beta}
\\
&=& \frac{1}{\beta^2}e^{-r^2/\beta} \prod\limits_{i=1}^n \Gamma_i
\left(m_i+\not\! v_i-\frac{1}{\beta}\not\! r
+\frac{1}{2}\not\!\partial_r\right). \nonumber
\end{eqnarray}
We thus have
\begin{eqnarray}
I_n(p_1,...,p_n) &=& \prod\limits_{i=1}^n
\int\limits_0^\infty\frac{d\beta_i}{\beta^2} \times
\exp\left\{-\sum\limits_{i=1}^{n}\beta_i m^2_i
+\frac{1}{\beta}\sum\limits_{1\le i<j\le 2\,n}
\beta_i\,\beta_j\,(v_i-v_j)^2\right\}
\\
&\times& \frac{1}{4}{\rm tr}\, \left[\prod\limits_{i=1}^n \Gamma_i
\left(m_i+\not\! v_i-\frac{1}{\beta}\not\! r
+\frac{1}{2}\not\!\partial_r\right)\right]\, , \nonumber
\end{eqnarray}
where $r=\sum\limits_{i=1}^{2\,n}\beta_i\,v_i$. Finally, we effect
some further transformations on the integration variables to get
the integral into a form  suitable for numerical evaluation:
\begin{itemize}
\item we use the formula
$$
\prod\limits_{i=1}^n\int\limits_0^\infty\!
d\beta_i\,f(\beta_1,...,\beta_n) = \int\limits_0^\infty\! dt\,
t^{n-1}\prod\limits_{i=1}^n\int\limits_0^\infty\! d\alpha_i\,
\delta\left(1-\sum\limits_{i=1}^n \alpha_i\right)
f(t\,\alpha_1,...,t\,\alpha_n)\, ;
$$
\item we scale the $t$ variable by $t\to w\,t$ with
      $w=\sum\limits_{i=1}^n \beta_{i+n}=\sum\limits_{i=1}^n\,s_{i}$
and introduce the new variable $\tilde \beta_i=\beta_{i}/w$
($i=n+1,...,2n)$.
\end{itemize}
We have
\begin{eqnarray}
I_n(p_1,...,p_n) &=& \int\limits_0^\infty\! dt
\frac{t^{n-1}}{(1+t)^2} \prod\limits_{i=1}^n
\int\limits_0^\infty\! d\alpha_i\,
\delta\left(1-\sum\limits_{i=1}^n \alpha_i\right)\cdot e^{-w\,z}
\label{master}
\\
&\times& \frac{w^2}{4}{\rm tr}\, \left[\prod\limits_{i=1}^n
\Gamma_i \left(m_i+\not\! v_i-\frac{1}{1+t}\not\! \tilde r
+\frac{1}{2\,w}\not\!\partial_{\tilde r}\right)\right]\, ,
\nonumber
\end{eqnarray}
where
$$
\tilde r = t\,\sum_{i=1}^n\alpha_i\,v_i
         +\sum_{i=n+1}^{2\,n}\tilde\beta_i\,v_i\, .
$$
The $z$ form in the exponential function is written as
\begin{eqnarray}
\label{z-form} z &=& t\,z_{\,\rm
loc}-\frac{t}{1+t}\,z_1-\frac{1}{1+t}\,z_2\, ,
\\
&&\nonumber\\
z_{\,\rm loc} &=&\sum_{i=1}^n\alpha_i\,m^2_i -\sum_{1\le i<j\le
n}\alpha_i\,\alpha_j\,a_{ij}\, ,
\nonumber\\
z_1
&=&\sum_{i=1}^n\alpha_i\,\sum_{j=n+1}^{2\,n}\tilde\beta_j\,a_{ij}
-\sum_{1\le i<j\le n}\alpha_i\,\alpha_j\,a_{ij}\, ,
\nonumber\\
z_2 &=&\sum_{n+1 \le i<j\le 2\,n}\tilde\beta_i\,
\tilde\beta_j\,a_{ij}\, . \nonumber
\end{eqnarray}
The matrix $a_{ij}=(v_i-v_j)^2$  depends on the invariant
kinematical variables. Explicit expressions for this matrix for
the triangle and box diagrams are given in Tables \ref{tab:trn}
and \ref{tab:box}, respectively. We will introduce the variable
$v=t/(1+t)$ ($0\le v \le 1$) in Eq.~(\ref{master}) in what
follows.

\begin{table}
\begin{center}
\def\arraystretch{2}
\begin{tabular}{|l|l|l|}
\hline
 $a_{12}=p_2^2$ & $a_{13}=p_1^2$ &  $a_{23}=p_3^2$ \\
\hline\hline
 $a_{14}=\left(c_{13}^1\right)^2 p_1^2$ &
 $a_{15}=\left(c_{12}^1\right)^2 p_2^2$ &
 $a_{16}=c_{23}^2\, p_1^2-c_{23}^2\, c_{23}^3\, p_3^2+c_{23}^3\, p_2^2$  \\
\hline
 $a_{24}=c_{13}^3\, p_2^2-c_{13}^1\, c_{13}^3\, p_1^2+c_{13}^1\, p_3^2$  &
 $a_{25}=\left(c_{12}^2\right)^2 p_2^2$ &
 $a_{26}=\left(c_{23}^2\right)^2 p_3^2$ \\
\hline
 $a_{34}=\left(c_{13}^3\right)^2 p_1^2$ &
 $a_{35}=c_{12}^2\, p_1^2-c_{12}^1\, c_{12}^2\, p_2^2+c_{12}^1\, p_3^2$  &
 $a_{36}=\left(c_{23}^3\right)^2 p_3^2$ \\
\hline
 $a_{45}=c_{13}^1\,(c_{13}^1-c_{12}^1)\, p_1^2$ &
 $a_{46}=c_{13}^3\,(c_{13}^3-c_{23}^3)\, p_1^2$ &
 $a_{56}=c_{12}^2\,(c_{12}^2-c_{23}^2)\, p_2^2$ \\
\hspace{0.5cm}        $+c_{12}^1\,(c_{12}^1-c_{13}^1)\, p_2^2
                     +c_{12}^1\,c_{13}^1\, p_3^2 $             &
\hspace{0.5cm}        $+c_{23}^3\,(c_{23}^3-c_{13}^3)\, p_2^2
        +c_{13}^3\,c_{23}^3\, p_3^2 $                          &
\hspace{0.5cm}        $+c_{23}^2\,(c_{23}^2-c_{12}^2)\, p_3^2
        +c_{12}^2\,c_{23}^2\, p_1^2 $ \\
\hline
\end{tabular}
\caption{\label{tab:trn}The matrix $a_{ij}=(v_i-v_j)^2$ for the
triangle diagram. Here, $v_1=0$, $v_2=-p_2$, $v_3=p_1$,
$v_4=c_{13}^1\, p_1$, $v_5=-c_{12}^1\, p_2$, $v_6=p_1+c_{23}^3\,
p_3$.}
\end{center}
\end{table}

\begin{table}
\begin{center}
\def\arraystretch{2}
\begin{tabular}{|l|l|l|}
\hline
 $a_{12}=p_2^2$ & $a_{13}=(p_2+p_3)^2$ &  $a_{14}=p_1^2$ \\
\hline
 $a_{23}=p_3^2$ & $a_{24}=(p_1+p_2)^2$ &  $a_{34}=p_4^2$ \\
\hline\hline
 $a_{15}=\left(c_{14}^1\right)^2 p_1^2$ &
 $a_{16}=\left(c_{12}^1\right)^2 p_2^2$ &
 $a_{17}=c_{23}^3\, p_2^2-c_{23}^2\, c_{23}^3\, p_3^2$ \\
 & & \hspace*{0.5cm} $ + c_{23}^2\, (p_2+p_3)^2$  \\
\hline
 $a_{18}=c_{34}^3\, p_1^2-c_{34}^3\, c_{34}^4\, p_4^2$ &
 $a_{25}=c_{14}^4 p_2^2-c_{14}^1\, c_{14}^4\, p_1^2$   &
 $a_{26}=\left(c_{12}^2\right)^2 p_2^2$ \\
       \hspace*{0.5 cm}  $+c_{34}^4\, (p_1+p_4)^2$  &
       \hspace*{0.5 cm}  $+c_{14}^1\, (p_1+p_2)^2$  & \\
\hline
 $a_{27}=\left(c_{23}^2\right)^2 p_3^2$                 &
 $a_{28}=c_{34}^4\, p_3^2-c_{34}^3\, c_{34}^4\, p_4^2 $ &
 $a_{35}=c_{14}^1\, p_4^2-c_{14}^1\, c_{14}^4\, p_1^2 $ \\
 & \hspace*{0.5 cm} $ +c_{34}^3\, (p_3+p_4)^2$ &
   \hspace*{0.5 cm} $ +c_{14}^4\, (p_1+p_4)^2$  \\
\hline
 $a_{36}=c_{12}^1\, p_3^2-c_{12}^1\, c_{12}^2\, p_2^2
        +c_{12}^2\, (p_2+p_3)^2$  &
 $a_{37}=\left(c_{23}^3\right)^2 p_3^2$ &
 $a_{38}=\left(c_{34}^3\right)^2 p_4^2$ \\
\hline
 $a_{45}=\left(c_{14}^4\right)^2 p_1^2$                 &
 $a_{46}=c_{12}^2\, p_1^2-c_{12}^1\, c_{12}^2\, p_2^2 $ &
 $a_{47}=c_{23}^2\, p_4^2-c_{23}^2\, c_{23}^3\, p_3^2$ \\
      &  \hspace*{0.5 cm} $ +c_{12}^1\, (p_1+p_2)^2$  &
         \hspace*{0.5 cm} $ +c_{23}^3\, (p_3+p_4)^2$ \\
\hline
 $a_{48}=\left(c_{34}^4\right)^2 p_4^2$ & & \\
\hline
\end{tabular}
\caption{\label{tab:box}The matrix $a_{ij}=(v_i-v_j)^2$ for the
box diagram. Here, $v_1=0$, $v_2=-p_2$, $v_3=-p_2-p_3$, $v_4=p_1$,
$v_5=c_{14}^1\, p_1$, $v_6=-c_{12}^1\, p_2$, $v_7=-p_2-c_{23}^2\,
p_3$, $v_8=p_1+c_{34}^4\, p_4$.}
\end{center}
\end{table}

Some further remarks are appropriate. One can see that the
existence of the loop integral Eq.~(\ref{master}) is defined by the
local $\alpha$ form $z_{\,\rm loc}$. Even if we apply the
constraints  Eq.~(\ref{conf}) it does not guarantee that this form
is always positive. For instance, in the simplest triangle diagram
with $p_1^2=p_2^2=p_3^2\equiv p^2$ and $m_1=m_2=m_3=m$ the form
$z_{\,\rm loc}=0$ if $p^2=3 m^2$. Such singularities in the
Feynman diagrams with local propagators are called anomalous
thresholds (see the discussion in Ref.~\cite{Davydychev:2001uj}
and other references therein). The situation is much more
complicated in the case of the box diagrams. Obviously, we cannot
consider the annihilation processes when the energy is large
enough to go beyond the normal thresholds corresponding to quark
production. However, in the case of the dissociation processes
considered here, the local form $z_{\,\rm loc}$ is always positive,
and we are able to make  self-consistent and reliable predictions
for physical observables.

\section{The $J/\psi$ dissociation amplitudes and cross sections}

\begin{figure}
\begin{center}
\epsfig{figure=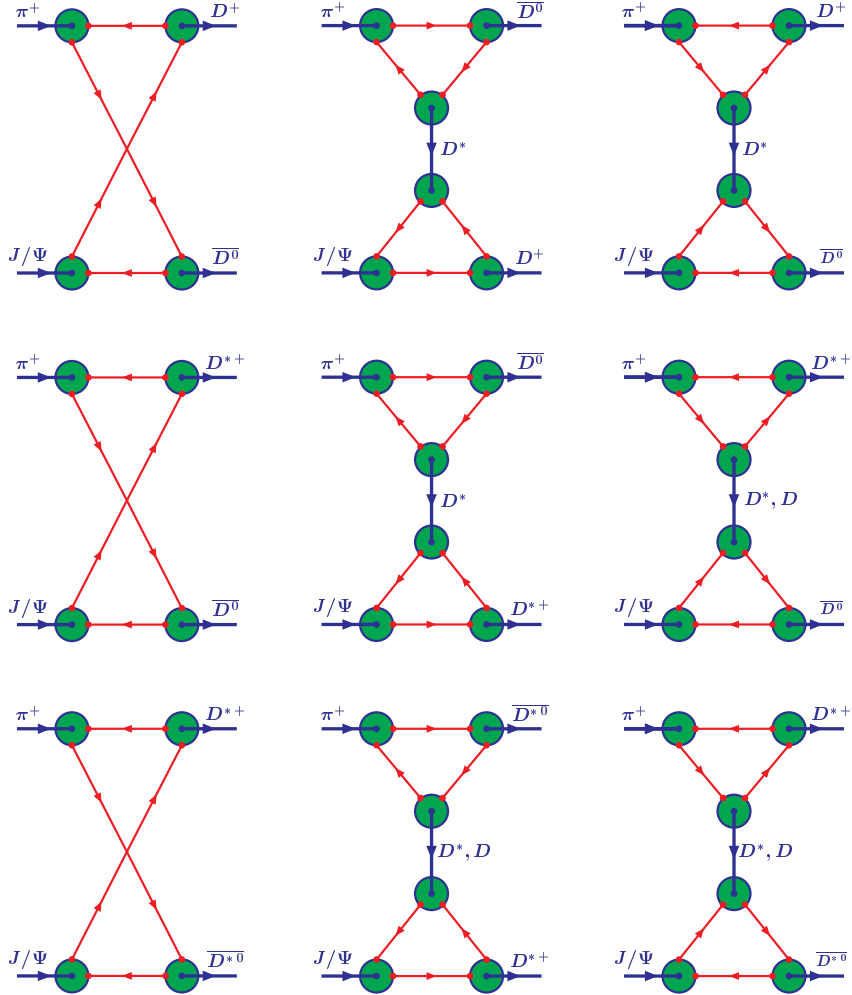,height=15cm}
\end{center}
\caption{ The Feynman diagrams describing the charm dissociation
processes
               $J/\psi+\pi^+\to D^+ +\overline {D^0}$,
               $D^{\ast\,+}+\overline{D^0}$,
               $D^{\ast\,+}+\overline {D^{\ast\,0}}$.}
\label{fig:JPDD}
\end{figure}

In our approach the dissociation processes
               $J/\psi+\pi^+\to D+\overline {D^0}$,
               $D^{\ast\,+}+\overline{D^0}$, and
               $D^{\ast\,+}+\overline {D^{\ast\,0}}$
are described by the diagrams in Fig.~\ref{fig:JPDD}.

Our momentum labeling is defined by
\begin{equation}
J/\psi(p_1)+\pi^+(p_2)\to D^+_3(q_1)+\overline{D_4^0}(q_2)
\end{equation}
where $D_3^+=D^+$ or $D^{\ast\,+}$,
      $\overline{D_4^0}=\overline{D^0}$ or $\overline{D^{\ast\,0}}$,
      $p_1^2=m_{H1}^2\equiv m_{J/\psi}^2$,
      $p_2^2=m_{H2}^2\equiv m_\pi^2$,
      $q_1^2=m_{H3}^2\equiv m^2_{D^+}\,( m^2_{D^{\ast\,+}})$,
      $q_2^2=m_{H4}^2\equiv m^2_{\overline{D^0}}\,(m^2_{\overline{D^{\ast\,0}}})$.

The Mandelstam variables are defined in the standard form
\begin{eqnarray*}
s &=& (p_1+p_2)^2=(q_1+q_2)^2,
\\
t &=& (p_1-q_1)^2=(p_2-q_2)^2,
\\
u &=& (p_1-q_2)^2=(p_2-q_1)^2,
\end{eqnarray*}
where $s+t+u=m_{H1}^2+m_{H2}^2+m_{H3}^2+m_{H4}^2$.

Cross sections are calculated by using the formula
\begin{equation}
\label{cross-sec} \sigma(s) = \frac{1}{192\pi
s}\frac{1}{p^2_{1,\rm cm}} \int\limits_{t_-}^{t_+}\! dt\,
|M(s,t)|^2
\end{equation}
where $M(s,t)$ is an invariant amplitude and
\begin{eqnarray*}
t_\pm &=& (E_{1,\rm cm}-E_{3,\rm cm})^2
         -(p_{1,\rm cm}\mp q_{1,\rm cm})^2,
\\
&&\\
E_{1,\rm cm}&=&\frac{s+m^2_{H1}-m^2_{H2}}{2\sqrt{s}}, \hspace{1cm}
E_{3,\rm cm}=\frac{s+m^2_{H3}-m^2_{H4}}{2\sqrt{s}},
\\
&&\\
p_{1,\rm
cm}&=&\frac{\lambda^{1/2}(s,m^2_{H1},m^2_{H2})}{2\sqrt{s}},
\hspace{1cm} q_{1,\rm
cm}=\frac{\lambda^{1/2}(s,m^2_{H3},m^2_{H4})}{2\sqrt{s}}.
\end{eqnarray*}
The reaction threshold is equal to $s_0=(m_{H3}+m_{H4})^2$. Note
that Eq.~(\ref{cross-sec}) contains the statistical factor 1/3
which comes from averaging over the initial state
$J/\psi$ polarizations.

The dissociation processes are described by both the box and the
resonance diagrams as shown in Fig.~\ref{fig:JPDD}. The resonance
diagrams depend explicitly only on the  $t$ or $u$ variables
whereas the box diagrams are functions of $s$ and $t$.

In the following three subsections we write explicitly the
amplitudes for the processes $ J/\psi + \pi^+ \to D^{+} +
\overline{D^{0}}$, $D^{\ast\,+} + \overline{D^{0}}$, $D^{\ast\,+}
+ \overline{D^{\ast\,0}}$ in terms of form factors; all the
analytical  expressions for them are reported.

\subsection{The channel
\boldmath{ $ J/\psi + \pi^+ \to D^{+} + \overline{D^{0}}$ }}

The invariant matrix element is written as%
\footnote{Note that the symbol $\varepsilon^{\alpha \beta p q}
\equiv \varepsilon^{\alpha \beta \mu \nu } p_\mu q_\nu$, where
$\varepsilon^{\alpha \beta \mu \nu }$ is the Levi-Civita tensor.
Moreover, the order of indices $\rm V$ and $\rm P$ of the form
factors should be read looking at the Fig.~\ref{fig:JPDD} and
following the arrows backward in the loops. }
\begin{equation}
\label{VPPP} M({\rm VPPP}) = M_\Box({\rm VPPP})\,+\, M_{\rm
res}({\rm VPPP}),
\end{equation}
where
\begin{eqnarray}
M_\Box({\rm VPPP}) &=& \epsilon_{\mu_1}(p_1)\cdot
\varepsilon^{\mu_1 p_1 p_2 q_2}\,F_{\rm VPPP}(s,t),
\\
M_{\rm res}({\rm VPPP}) &=& \epsilon_{\mu_1}(p_1)\cdot \left\{ -\,
M^{\alpha}_{\triangle\,a}({\rm PVP})\,
D^{\alpha\beta}_{D^\ast}(p_2-q_2)\,
M^{\mu_1\beta}_{\triangle\,a}({\rm VVP}) \right.
\nonumber\\
&-& \left. M^{\alpha}_{\triangle\,b}({\rm PPV})\,
D^{\alpha\beta}_{D^\ast}(p_2-q_1)\,
M^{\mu_1\beta}_{\triangle,b}({\rm VPV}) \right\}\,,
\nonumber\\
&&\nonumber\\
M^{\alpha}_{\triangle\,a}({\rm PVP}) &=& p_2^\alpha\,F^{(1a)}_{\rm
PVP}(t), \hspace{1cm} M^{\mu_1\beta}_{\triangle\,a}({\rm VVP}) =
\varepsilon^{\mu_1\beta p_1(p_2-q_2)}\,F^{(a)}_{\rm VVP}(t),
\nonumber\\
&&\nonumber\\
M^{\alpha}_{\triangle\,b}({\rm PPV}) &=& p_2^\alpha\,F^{(1b)}_{\rm
PPV}(u), \hspace{1cm} M^{\mu_1\beta}_{\triangle\,b}({\rm VPV}) =
\varepsilon^{\mu_1\beta p_1(p_2-q_1)}\,F^{(b)}_{\rm VPV}(u).
\nonumber
\end{eqnarray}
The common minus sign is due to the extra quark loop in the
resonance diagrams as compared to the box diagram. This extra sign
is very important, as will be seen later on, since the extra sign
leads to a constructive interference of resonance and box graph
contributions. The loop momenta directions come from explicit
calculation of the S-matrix elements and are shown in
Fig.~\ref{fig:JPDD}.
The vector ($D^\ast$) and pseudoscalar ($D$) meson propagators are
given by

$$
D^{\alpha\beta}_{D^\ast}(p) = \frac{-g^{\alpha\beta}+p^\alpha
p^\beta/m^2_{D^\ast}}
     {m^2_{D^\ast}-p^2},
\hspace{1cm} i^2 D_{D}(p) =\frac{-1}{m^2_{D}-p^2}.
$$

For the form factors one obtains
\begin{eqnarray*}
F_{\rm VPPP}(s,t) &=& g_{J/\psi}\,g^2_{D}\,g_\pi \int\!
d\sigma_\Box \cdot\exp(-w\cdot z_\Box(s,t))\cdot w^2\,(
m_c\,(1+d_2 )- m_q\,d_2),
\\
F^{(1a)}_{\rm PVP}(t) &=& g_\pi\,g_{D^\ast}\,g_D\,\int\!
d\sigma_\triangle \cdot \exp(-w z_\triangle(t))\, \cdot
[w\,(-m_q\,m_c-m_q^2\,b_1
\\
&& +m^2_\pi\,b_1^2\,(b_1+b_2-1)-m^2_D\,b_1^2\,b_2
\\
&& +t\,b_2\,(b_1\,b_2+b_1^2-1+b_2))
\\
&& -(1-v)\,(1+3\,b_1)],
\\
&&\\
F^{(a)}_{\rm VVP}(t) &=& g_{J/\psi}\,g_{D^\ast}\,g_D\,\int\!
d\sigma_\triangle \cdot \exp(-w z_\triangle(t))\, \cdot[w\,
(m_c\,(b_2 - 1) - m_q\,b_2 )],
\\
&&\\
F^{(1b)}_{\rm PPV}(u) &=& g_{\pi}\,g_{D^\ast}\,g_D\,\int
d\sigma_\triangle \cdot \exp(-w z_\triangle(u))\, \cdot [ w\,(
m_q\,m_c  + m_q^2 \, ( 1 - b_1 - b_2 )
\\
&& + m^2_\pi \, (  - 2\,b_1\,b_2 + b_1\,b_2^2 + b_1 +
2\,b_1^2\,b_2 - 2\,b_1^2 + b_1^3 )
\\
&& + m^2_D \, (  - 2\,b_1\,b_2 + 2\,b_1\,b_2^2 + b_1^2\,b_2 + b_2
- 2\,b_2^2 + b_2^3 )
\\
&& + u \, ( 2\,b_1\,b_2 - b_1\,b_2^2 - b_1^2\,b_2 ))
\\
&&
 + 4 + 3\,b_1\,v - 3\,b_1 + 3\,b_2\,v - 3\,b_2 - 4\,v],
\\
F^{(b)}_{\rm VPV}(u) &=& g_{J/\psi}\,g_D\,g_{D^\ast}\,\int\!
d\sigma_\triangle \cdot \exp(-w z_\triangle(u)) \cdot
[w\,(-m_c+b_2\,(m_c-m_q))].
\end{eqnarray*}

The integration measures are defined by
\begin{eqnarray}
\int\! d\sigma_\Box &=&\frac{3}{4\pi^2}\, \int\limits_0^1 dv
\left(\frac{v}{1-v}\right)^3 \int
d^4\alpha\,\delta\left(1-\sum\limits_{i=1}^4\alpha_i\right),
\nonumber\\
&&\\
\int d\sigma_\triangle &=& \frac{3}{4\pi^2}\, \int\limits_0^1\! dv
\left(\frac{v}{1-v}\right)^2 \int
d^3\alpha\,\delta\left(1-\sum\limits_{i=1}^3\alpha_i\right).
\nonumber
\end{eqnarray}

The coefficients $b_i$ (triangle diagrams) and $d_i$ (box
diagrams) are given by
\begin{eqnarray*}
b_1 &=& v\,\alpha_3+(1-v)\,(\tilde s_1\, c_{13}^1+\tilde s_3\,
c_{23}^2),
\\
b_2 &=& v\,\alpha_2+(1-v)\,(\tilde s_2\, c_{12}^1+\tilde s_3\,
c_{23}^3),
\\
&&\\
d_1 &=& v\,\alpha_4+(1-v)\,(\tilde s_1\, c_{14}^1+\tilde s_4\,
c_{34}^3),
\\
d_2 &=& -\,v\,(\alpha_2+\alpha_3)
        -(1-v)\,(\tilde s_2\, c_{12}^1+\tilde s_3+\tilde s_4\, c_{34}^4),
\\
d_3 &=& -\,v\,\alpha_3
        -(1-v)\,(\tilde s_3\, c_{23}^2+\tilde s_3+\tilde s_4\, c_{34}^4)\,.
\end{eqnarray*}

The values of $z_\triangle$, $z_\Box$, and $w$ are defined by the
relevant diagrams (size parameters and quark and hadron masses).

\subsection{The channel
\boldmath{$J/\psi+\pi^+\to D^{\ast\,+} + \overline{D^{0}}$}}

The invariant matrix element can be written as:
\begin{equation}
\label{VPPV} M({\rm VPPV}) = M_\Box({\rm VPPV})\,+\, M_{\rm
res}({\rm VPPV})\, ,
\end{equation}
with
\begin{eqnarray}
M_\Box({\rm VPPV}) &=&
\epsilon_{\mu_1}(p_1)\epsilon_{\mu_2}(q_1)\cdot
\left(p_2^{\mu_1}\,p_1^{\mu_2}\,F^{(1)}_{\rm VPPV}(s,t)
     +q_2^{\mu_1}\,p_1^{\mu_2}\,F^{(2)}_{\rm VPPV}(s,t)
\right.
\nonumber\\
&+& \left.
   p_2^{\mu_1}\,p_2^{\mu_2}\,F^{(3)}_{\rm VPPV}(s,t)
     +q_2^{\mu_1}\,p_2^{\mu_2}\,F^{(4)}_{\rm VPPV}(s,t)
     +g^{\mu_1\mu_2}\,F^{(5)}_{\rm VPPV}(s,t)
\right)\, ,
\nonumber\\
&&\nonumber\\
M_{\rm res}({\rm VPPV}) &=&
\epsilon_{\mu_1}(p_1)\epsilon_{\mu_2}(q_1)
\nonumber\\
&\times& \left( -\, M^{\alpha}_{\triangle\,a}({\rm PVP})\,
D^{\alpha\beta}_{D^\ast}(p_2-q_2)\,
M^{\mu_1\beta\mu_2}_{\triangle\,a}({\rm VVV}) -\,
M^{\alpha\mu_2}_{\triangle\,b}({\rm PVV})\,
D^{\alpha\beta}_{D^\ast}(p_2-q_1)\,
M^{\mu_1\beta}_{\triangle,b}({\rm VPV}) \right.
\nonumber\\
&+& \left. M^{\mu_2}_{\triangle\,c}({\rm PVP})\, D_{D}(p_2-q_1)\,
M^{\mu_1}_{\triangle,c}({\rm VPP}) \right)\,. \nonumber
\end{eqnarray}
The expressions for the form factors are given in the Appendix.

\subsection{The channel
\boldmath{$J/\psi+\pi^+\to D^{\ast\,+} + \overline{D^{\ast\,0}}$}}

The amplitude for the process $J/\psi+\pi^+\to D^{\ast\,+} +
\overline{D^{\ast\,0}}$ can be written as
\begin{equation}
\label{VVPV} M({\rm VVPV}) = M_\Box({\rm VVPV})\,+\, M_{\rm
res}({\rm VVPV}),
\end{equation}
where
\begin{eqnarray}
M_\Box({\rm VVPV}) &=&
\epsilon_{\mu_1}(p_1)\epsilon_{\mu_2}(q_2)\epsilon_{\mu_3}(q_1)
\nonumber\\
&\times& \left(
 g^{\mu_2\mu_3}\varepsilon^{p_1 p_2 q_2 \mu_1}\,F^{(1)}_{\rm VVPV}(s,t)
+g^{\mu_1\mu_3}\varepsilon^{p_1 p_2 q_2 \mu_2}\,F^{(2)}_{\rm
VVPV}(s,t) +g^{\mu_1\mu_2}\varepsilon^{p_1 p_2 q_2
\mu_3}\,F^{(3)}_{\rm VVPV}(s,t) \right.
\nonumber\\
&& +p_1^{\mu_3}\varepsilon^{p_1 p_2 \mu_1 \mu_2}\,F^{(4)}_{\rm
VVPV}(s,t) +p_1^{\mu_2}\varepsilon^{p_1 p_2 \mu_1
\mu_3}\,F^{(5)}_{\rm VVPV}(s,t) +p_2^{\mu_3}\varepsilon^{p_1 p_2
\mu_1 \mu_2}\,F^{(6)}_{\rm VVPV}(s,t)
\nonumber\\
&& +p_2^{\mu_2}\varepsilon^{p_1 p_2 \mu_1 \mu_3}\,F^{(7)}_{\rm
VVPV}(s,t) +p_2^{\mu_1}\varepsilon^{p_1 p_2 \mu_2
\mu_3}\,F^{(8)}_{\rm VVPV}(s,t) +q_2^{\mu_1}\varepsilon^{p_1 p_2
\mu_2 \mu_3}\,F^{(9)}_{\rm VVPV}(s,t)
\nonumber\\
&& +p_2^{\mu_2}\varepsilon^{p_1 q_2 \mu_1 \mu_3}\,F^{(10)}_{\rm
VVPV}(s,t) +p_2^{\mu_3}\varepsilon^{p_1 q_2 \mu_1
\mu_2}\,F^{(11)}_{\rm VVPV}(s,t) +p_1^{\mu_3}\varepsilon^{p_2 q_2
\mu_1 \mu_2}\,F^{(12)}_{\rm VVPV}(s,t)
\nonumber\\
&& +p_1^{\mu_2}\varepsilon^{p_2 q_2 \mu_1 \mu_3}\,F^{(13)}_{\rm
VVPV}(s,t) +p_2^{\mu_1}\varepsilon^{p_2 q_2 \mu_2
\mu_3}\,F^{(14)}_{\rm VVPV}(s,t) +q_2^{\mu_1}\varepsilon^{p_2 q_2
\mu_2 \mu_3}\,F^{(15)}_{\rm VVPV}(s,t)
\nonumber\\
&& \left. +\varepsilon^{p_1 \mu_1 \mu_2 \mu_3}\,F^{(16)}_{\rm
VVPV}(s,t) +\varepsilon^{p_2 \mu_1 \mu_2 \mu_3}\,F^{(17)}_{\rm
VVPV}(s,t) \right). \nonumber
\end{eqnarray}

When taking into account parity invariance one knows from helicity
counting that there are only 14 independent amplitudes in the
$(VVPV)$ case. The above set of 17 amplitudes are in fact not
independent. They can be reduced to a set of 14 independent
amplitudes by making use of the Schouten identity (see, e.g.,
\cite{Korner:2003zq})

$$
g^{\mu\mu_1}\varepsilon^{\mu_2\mu_3\mu_4\mu_5} +{\rm
cycl.(\mu_1,\mu_2,\mu_3,\mu_4,\mu_5)}=0.
$$
We have made use of the Schouten identity as an additional check
on our numerical calculations.

Moreover, the resonance term, $M_{\rm res}({\rm VVPV})$, and all
the structures involved are reported in the following expressions:
\begin{eqnarray}
M_{\rm res}({\rm VVPV}) &=&
\epsilon_{\mu_1}(p_1)\epsilon_{\mu_2}(q_2)\epsilon_{\mu_3}(q_1)
\nonumber\\
&\times& \left( -\, M^{\alpha\mu_2}_{\triangle\,a}({\rm PVV})\,
D^{\alpha\beta}_{D^\ast}(p_2-q_2)\,
M^{\mu_1\beta\mu_3}_{\triangle\,a}({\rm VVV}) -\,
M^{\alpha\mu_3}_{\triangle\,b}({\rm PVV})\,
D^{\alpha\beta}_{D^\ast}(p_2-q_1)\,
M^{\mu_1\beta\mu_2}_{\triangle,b}({\rm VVV}) \right.
\nonumber\\
&+& \left. M^{\mu_2}_{\triangle\,c}({\rm PPV})\, D_{D}(p_2-q_2)\,
M^{\mu_1\mu_3}_{\triangle,c}({\rm VPV}) +\,
M^{\mu_3}_{\triangle,d}({\rm PVP})\, D_{D}(p_2-q_1)\,
M^{\mu_1\mu_2}_{\triangle,d}({\rm VVP}) \right)\,,
\nonumber\\
&&\nonumber\\
M^{\alpha\mu_2}_{\triangle\,a}({\rm PVV}) &=&
\varepsilon^{p_2(p_2-q_2)\mu_2\alpha}\,F^{(a)}_{\rm PVV}(t)\,,
\nonumber\\
&&\nonumber\\
&&\nonumber\\
M^{\mu_1\beta\mu_3}_{\triangle\,a}({\rm VVV}) &=&
 (p_2-q_2)^{\mu_1}p_1^\beta p_1^{\mu_3}\,F^{(1a)}_{\rm VVV}(t)
+(p_2-q_2)^{\mu_1}(p_2-q_2)^{\beta} p_1^{\mu_3}\,F^{(2a)}_{\rm
VVV}(t)
\nonumber\\
&+&
 g^{\mu_1\mu_3} p_1^{\beta}\,F^{(3a)}_{\rm VVV}(t)
+g^{\mu_1\mu_3} (p_2-q_2)^{\beta}\,F^{(4a)}_{\rm VVV}(t)
\nonumber\\
&+&
 g^{\mu_1\beta} p_1^{\mu_3}\,F^{(5a)}_{\rm VVV}(t)
+g^{\mu_3\beta}(p_2-q_2)^{\mu_1}\,F^{(6a)}_{\rm VVV}(t)\,,
\nonumber\\
&&\nonumber\\
&&\nonumber\\
M^{\alpha\mu_3}_{\triangle\,b}({\rm PVV}) &=&
\varepsilon^{p_2(p_2-q_1)\mu_3\alpha}\,F^{(b)}_{\rm PVV}(u)\,,
\nonumber\\
&&\nonumber\\
M^{\mu_1\beta\mu_2}_{\triangle\,b}({\rm VVV}) &=&
 (p_2-q_1)^{\mu_1}p_1^\beta p_1^{\mu_2}\,F^{(1b)}_{\rm VVV}(u)
+(p_2-q_1)^{\mu_1}(p_2-q_1)^{\beta} p_1^{\mu_2}\,F^{(2b)}_{\rm
VVV}(u)
\nonumber\\
&+&
 g^{\mu_1\mu_2} p_1^{\beta}\,F^{(3b)}_{\rm VVV}(u)
+g^{\mu_1\mu_2} (p_2-q_1)^{\beta}\,F^{(4b)}_{\rm VVV}(u)
\nonumber\\
&+&
 g^{\mu_1\beta} p_1^{\mu_2}\,F^{(5b)}_{\rm VVV}(u)
+g^{\mu_2\beta}(p_2-q_1)^{\mu_1}\,F^{(6b)}_{\rm VVV}(u)\,,
\nonumber\\
&&\nonumber\\
&&\nonumber\\
M^{\mu_2}_{\triangle\,c}({\rm PPV}) &=& p_2^{\mu_2}\,F^{(c)}_{\rm
PPV}(t)\,,
\nonumber\\
M^{\mu_1\mu_3}_{\triangle,c}({\rm VPV}) &=&
\varepsilon^{p_1q_1\mu_1\mu_3}\,F^{(c)}_{\rm VPV}(t)\,,
\nonumber\\
&&\nonumber\\
M^{\mu_3}_{\triangle,d}({\rm PVP})\, &=& p_2^{\mu_3}\,F^{(d)}_{\rm
PVP}(u)
\nonumber\\
M^{\mu_1\mu_2}_{\triangle,d}({\rm VVP}) &=&
\varepsilon^{p_1q_2\mu_1\mu_2}\,F^{(d)}_{\rm VVP}(u). \nonumber
\end{eqnarray}
All the expressions for the form factors are reported in the
Appendix.\\
Note that  the $J/\psi$ dissociation amplitudes
are not equal to zero when contracted with the four-momentum
of the $J/\psi$ except in Eq.~(\ref{VPPP}) which involves
the Levi-Civita tensor. In our approach we consider
the  $J/\psi$ and other vector mesons as  bound states
of constituent quarks and not as  gauge fields.

\section{Numerical results and discussions}
\label{s:results}

\begin{figure}
\begin{center}
\epsfig{figure=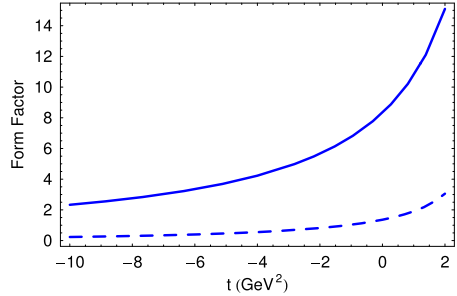,height=4cm}
\end{center}
\caption{Dependence of $F^{(1a)}_{\rm \pi D^\ast D}(t)$ (solid
line) and $F^{(a)}_{\rm J/\psi D^\ast D}(t)$ (dashed line) on the
invariant variable  `t'' which is the $D^\ast$-momentum squared.}
\label{fig:PVP-VVP}
\end{figure}

\begin{figure}
\begin{center}
\begin{tabular}{cc}
\epsfig{figure=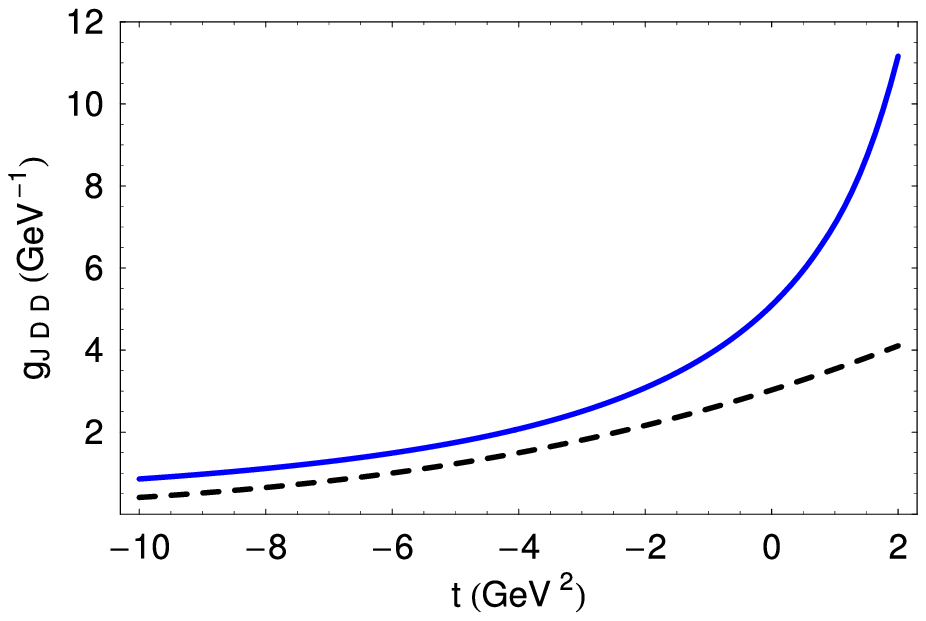,height=4cm} &
\epsfig{figure=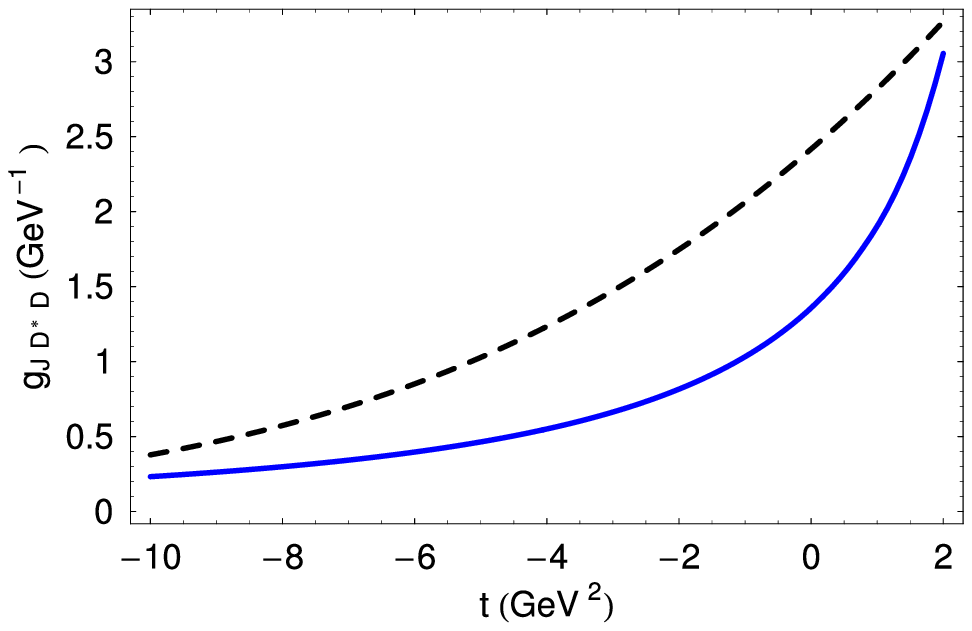,height=4cm}
\end{tabular}
\end{center}
\caption{Comparison of our (solid line) form factors
$F^{(1a)}_{\rm D\, J/\psi\, D}(t)$ and $F^{(a)}_{\rm J/\psi D^\ast
D}(t)$ with those (dashed line) obtained in
\cite{Matheus:2003pk}.}
\label{fig:comparison}
\end{figure}

Some comments about and comparisons of the $t$ dependence of the form
factors are in order. The behavior of $F^{(1a)}_{\rm PVP}(t)\,
=\, F^{(1a)}_{\rm \pi D^\ast D}(t)$  (in the literature, for $t\ =\
m_{D^\ast}^2$, it is called $g_{\rm D^\ast D \pi}$) and
$F^{(a)}_{\rm VVP}(t)\, =\, F^{(a)}_{\rm J/\psi\, D^\ast\, D}(t)$
in the kinematical region is shown in Figs.~\ref{fig:PVP-VVP} and ~\ref{fig:comparison}. In
order to be able to compare with other calculations we quote the
value  of $F^{(1a)}_{\rm \pi D^\ast D}(m_{D^\ast}^2)$ ($\equiv
g_{\rm D^\ast D \pi}$) which is equal to 22. This value is about
$2 \sigma$ larger than the recent experimental results from CLEO,
$g_{\rm D^\ast D \pi}\, =\,  17.9\pm 0.3 \pm
1.9$~\cite{CLEO_gDstDpi}. A very small value for $g_{\rm D^\ast D
\pi}$ was predicted by the light cone QCD sum rules approach
\cite{LCSR}.
\\
For the $F^{(a)}_{\rm J/\psi D^\ast \pi}(t)$ form factor, we
cannot go on the mass shell due to the presence of an anomalous
threshold.

\begin{figure}
\begin{center}
\begin{tabular}{cc}
\epsfig{figure=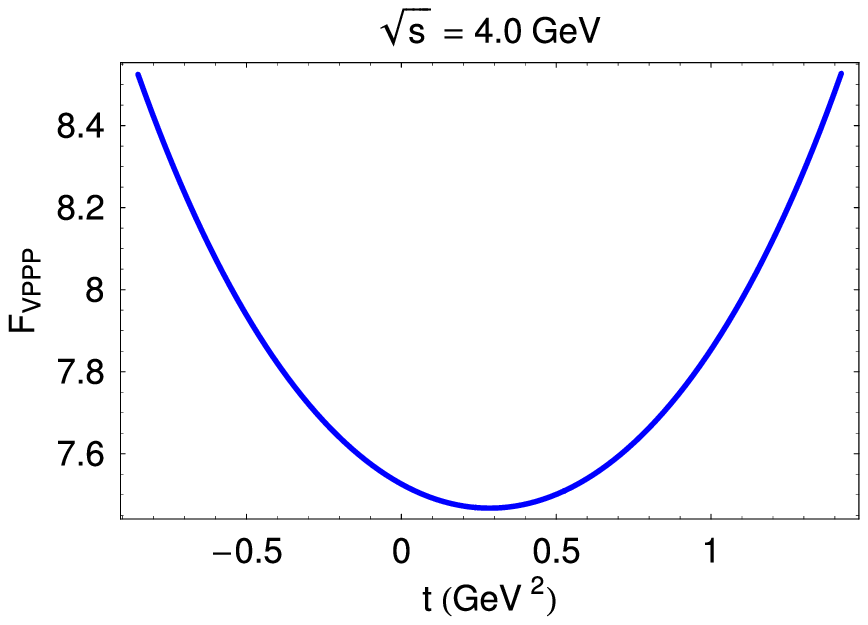,height=4cm} &
\epsfig{figure=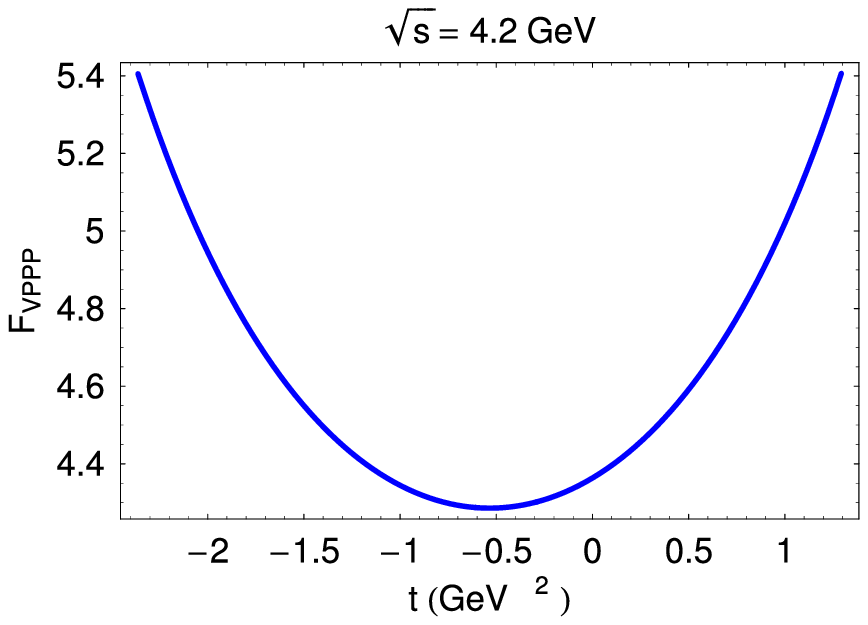,height=4cm} \\
\epsfig{figure=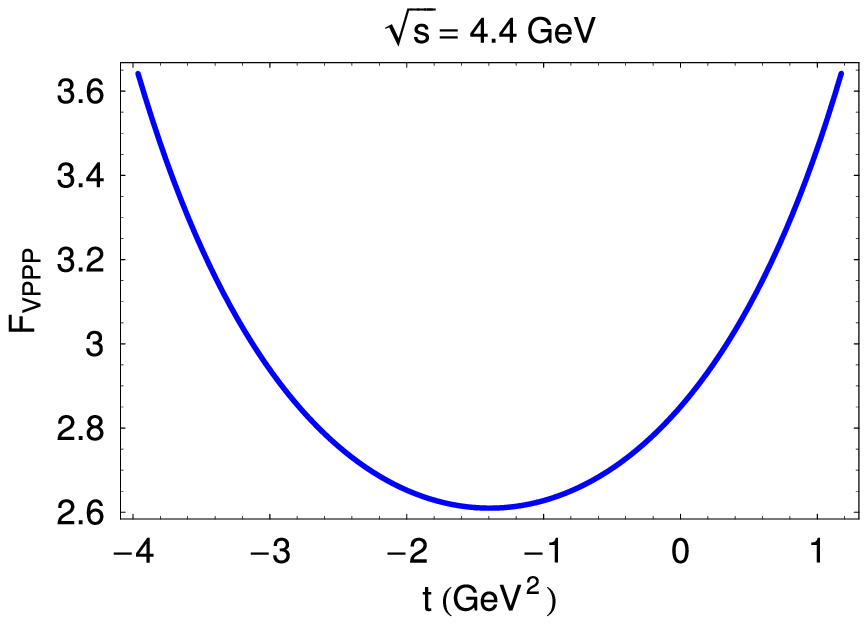,height=4cm} &
\epsfig{figure=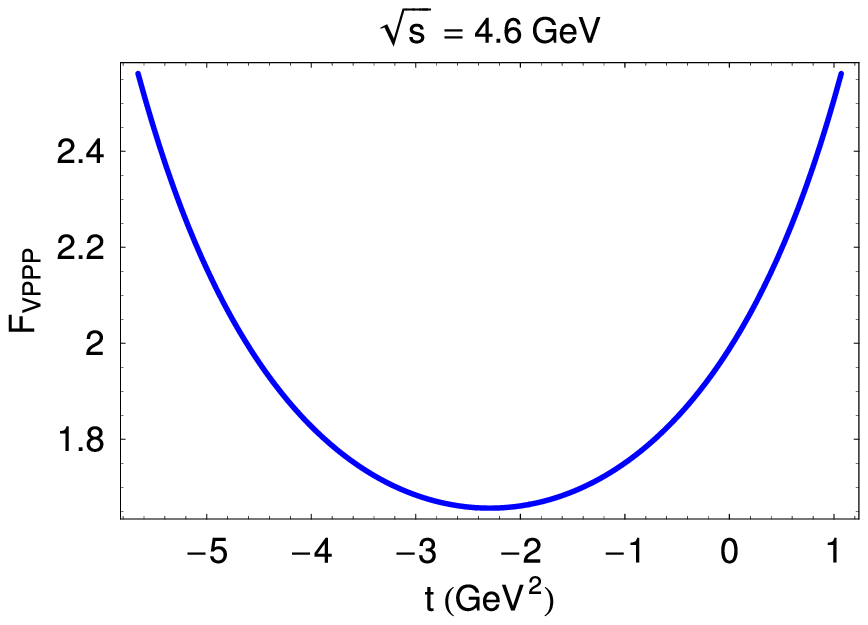,height=4cm} \\
\epsfig{figure=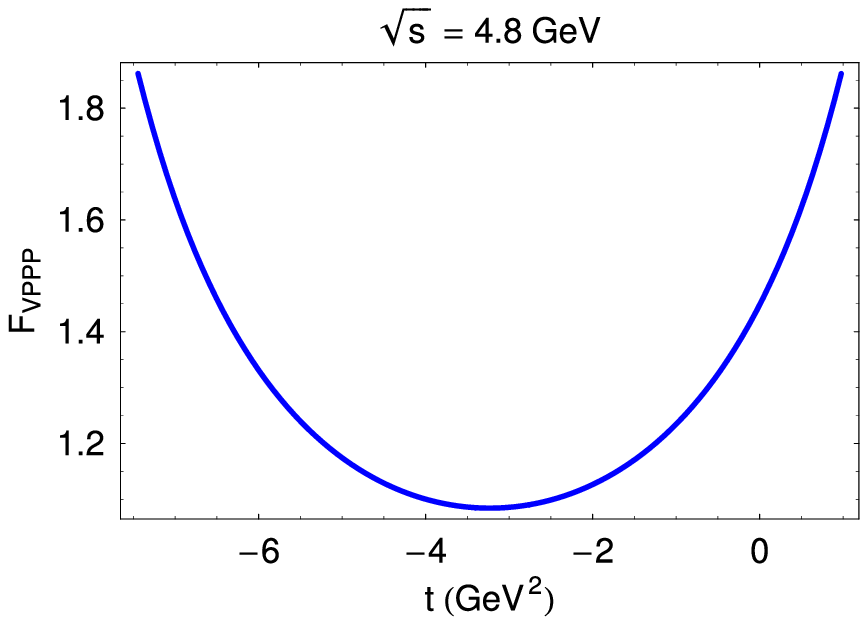,height=4cm} &
\epsfig{figure=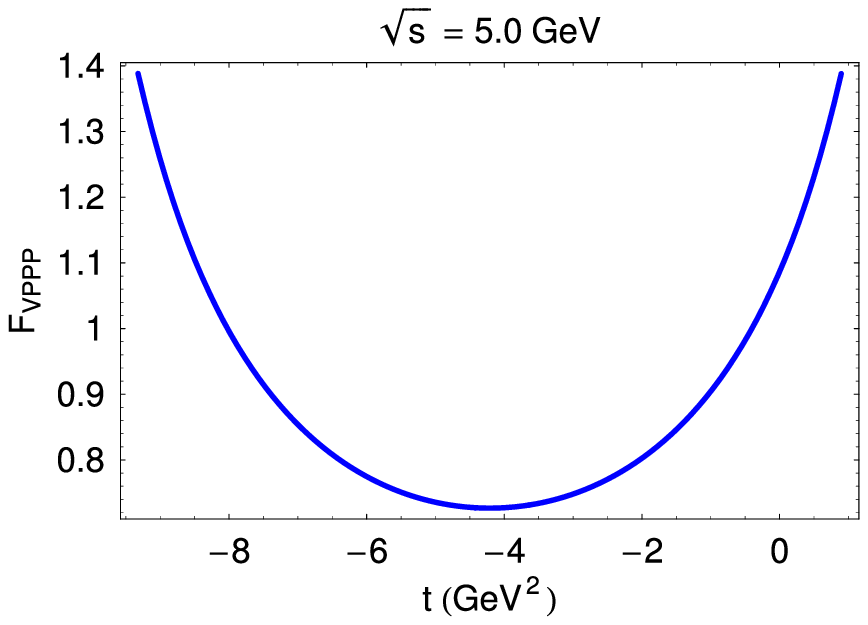,height=4cm} \\
\end{tabular}
\end{center}
\caption{Dependence of the form factor $F_{\rm J/\psi\, \bar D\,
\pi\, D}(s,t)$ on t for different values of $\sqrt{s}$. }
\label{fig:VPPP}
\end{figure}

\begin{figure}
\begin{center}
\epsfig{figure=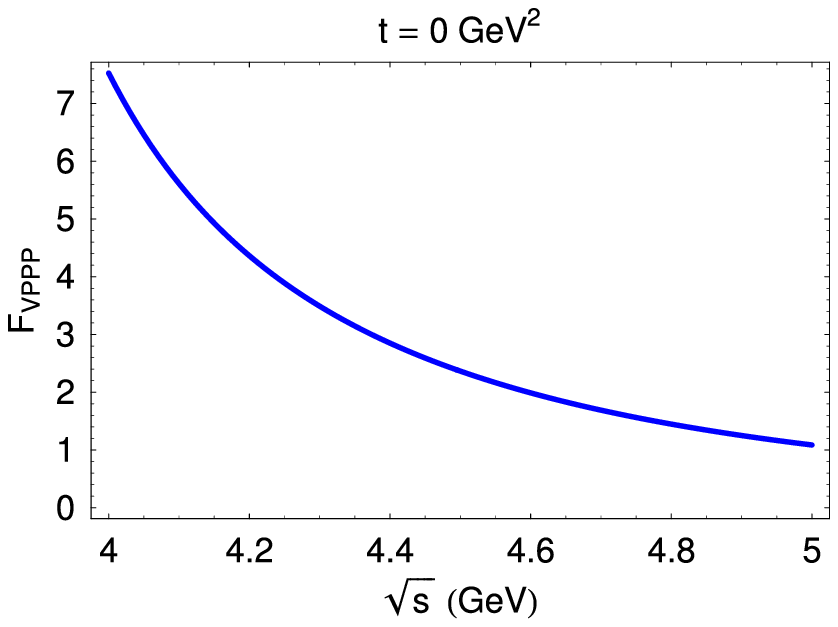,height=5cm}
\end{center}
\caption{Dependence of the form factor $F_{\rm J/\psi\, \bar D\,
\pi\, D}(s,t)$ on $s$ at $t=0$. } \label{fig:VPPP_s}
\end{figure}

The dependence of $F_{\rm J/\psi\, \bar D\, \pi\, D}(s,t)$ on $t$
for different values of $\sqrt{s}$ is shown in
Fig.~\ref{fig:VPPP}. One can see that the $t$ behavior is rather
flat. Moreover, the dependence of $F_{\rm J/\psi\, \bar D\, \pi\,
D}(s,t)$ on $\sqrt{s}$ at $t=0$ is shown in Fig.~\ref{fig:VPPP_s}.

\begin{figure}
\begin{center}
\epsfig{figure=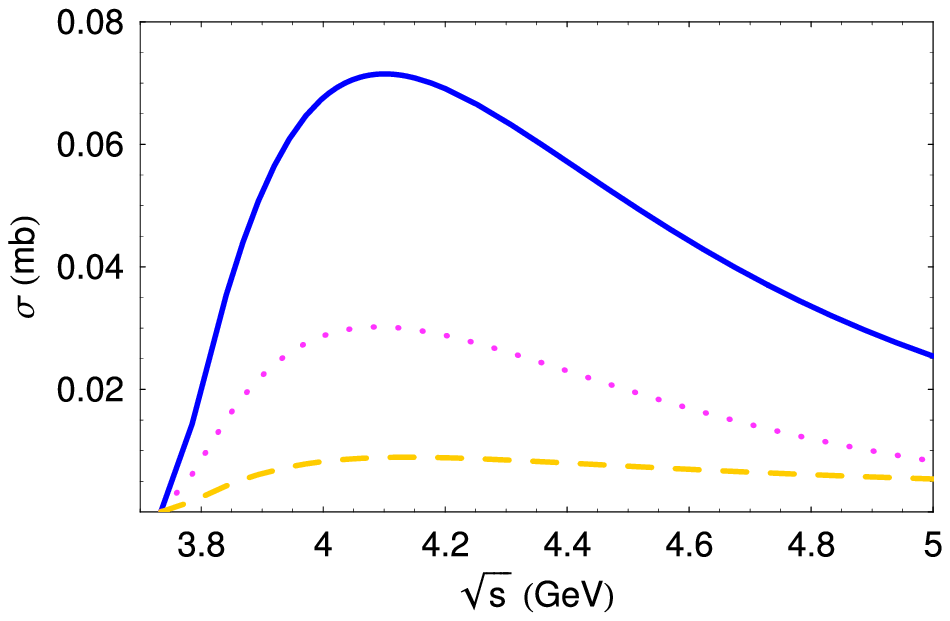,height=7cm}
\end{center}
\caption{\label{fig:JDDP}Contributions of the box (dotted curve)
and resonance (dashed curve) diagrams to total (solid curve) cross
section for the process $J/\psi+\pi^+\to D^+ + \overline{D^0}$.}
\end{figure}

\begin{figure}
\begin{center}
\begin{tabular}{cc}
\epsfig{figure=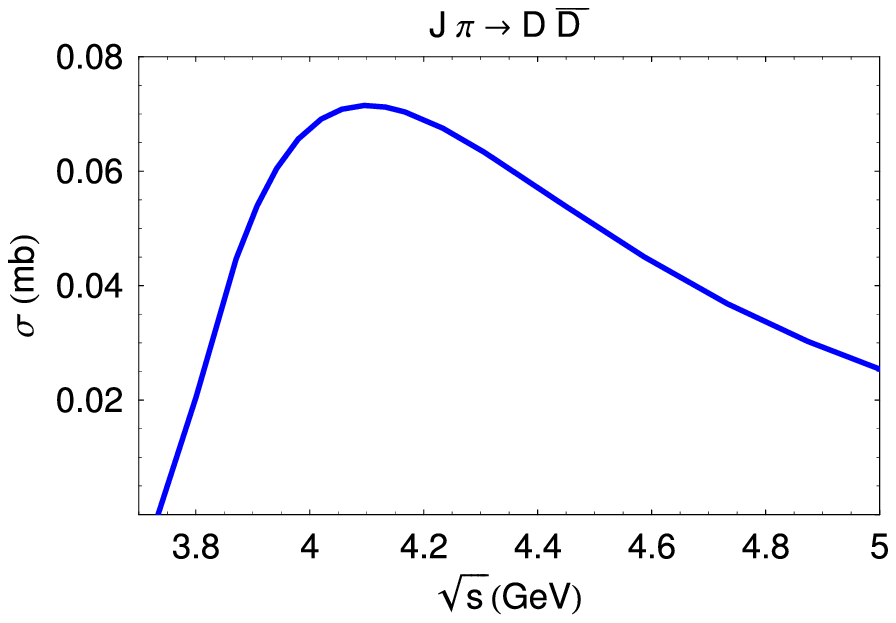,height=4truecm} &
\epsfig{figure=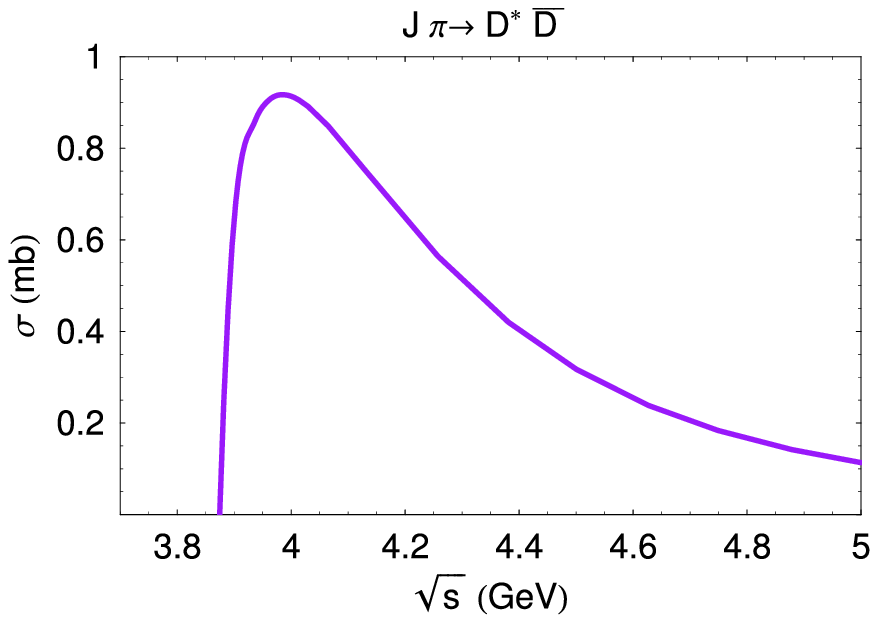,height=4truecm} \\
\epsfig{figure=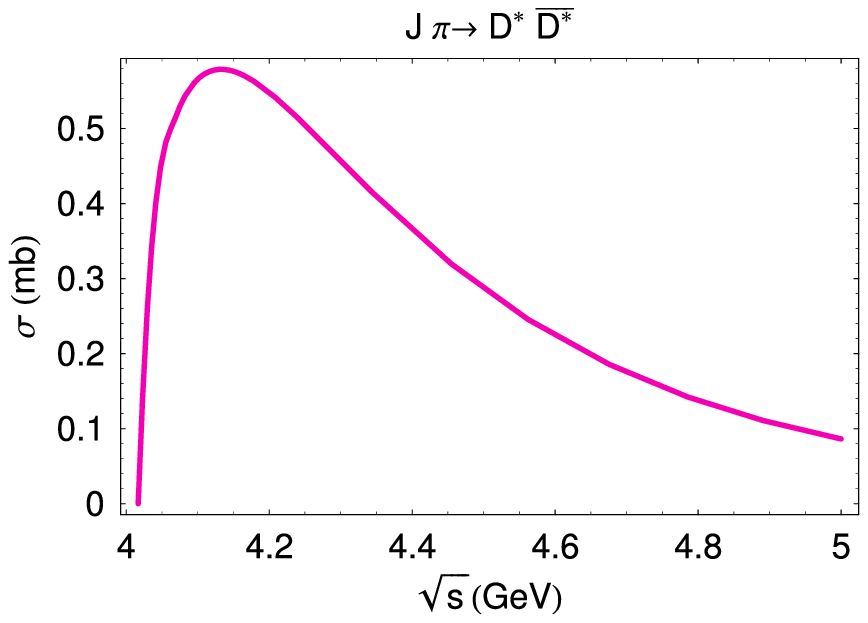,height=4truecm} &
\epsfig{figure=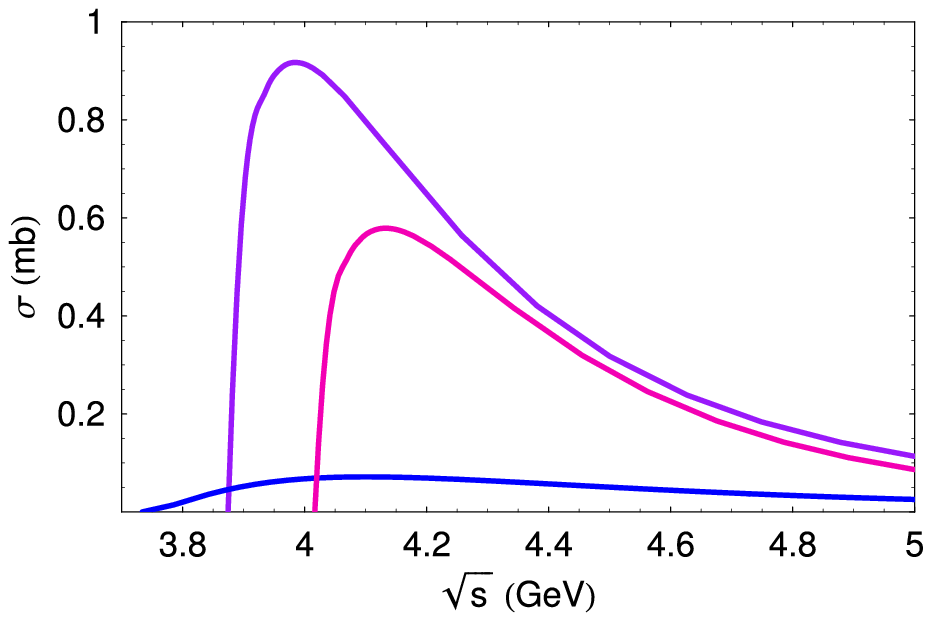,height=4truecm} \\
\end{tabular}
\end{center}
\caption{The cross sections of the processes $J/\psi+\pi^+\to D^+
+ \overline{D^0}$, $J/\psi+\pi^+\to D^{\ast\,+} + \overline{D^0}$
and $J/\psi+\pi^+\to D^{\ast\,+} + \overline{D^{\ast\,0}}$. All
three curves are shown together in the last plot.}
\label{fig:crossec}
\end{figure}

\begin{figure}
\begin{center}
\epsfig{figure=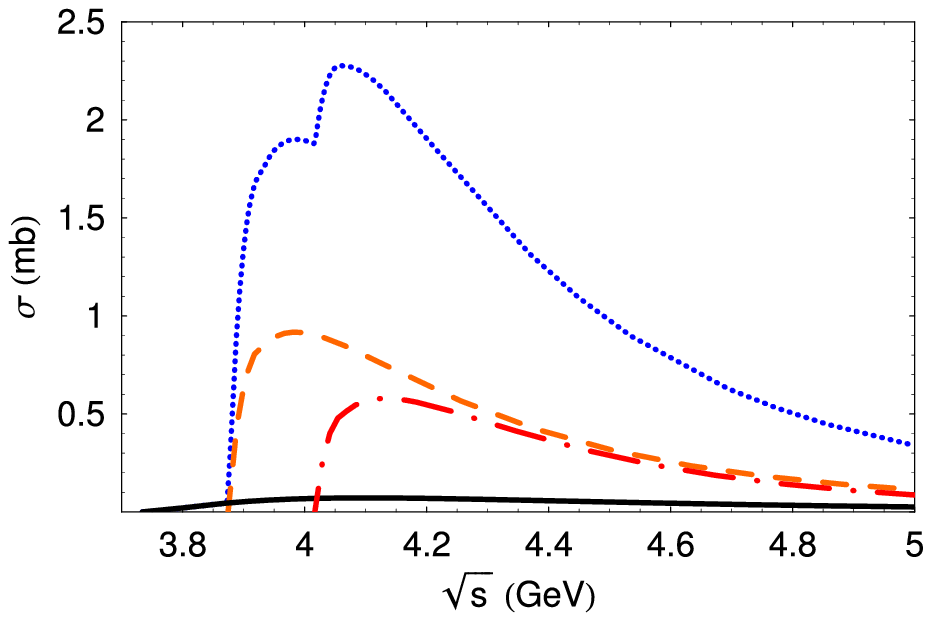,height=7cm}
\end{center}
\caption{The total cross section (dotted line) together with the
contributions already plotted in Fig.~\protect{\ref{fig:crossec}}}
\label{fig:sigmatot}
\end{figure}

In Fig.~\ref{fig:JDDP} we separately plot the contributions coming
from the box and resonance diagrams for the process
$J/\psi+\pi^+\to D^{+} + \overline{D^0}$.
We display  the dependence of the cross sections on the variable
$\sqrt{s}$  for each channel in Fig.~\ref{fig:crossec}. The total
cross section is a sum over all channels,
\begin{equation}
\label{total} \sigma_{\rm tot}(s)= \sigma_{D^{+}\overline{D^0}}(s)
+\sigma_{D^{\ast\,+}\overline{D^0}}(s)
+\sigma_{D^{+}\overline{D^{\ast\,0}}}(s)
+\sigma_{D^{\ast\,+}\overline{D^{\ast\,0}}}(s).
\end{equation}
which are plotted in Fig.~\ref{fig:crossec} Note that
$\sigma_{D^{+}\overline{D^{\ast\,0}}}
=\sigma_{D^{\ast\,+}\overline{D^0}}$. We plot $\sigma_{\rm
tot}(s)$ as a function of $\sqrt{s}$ in Fig.~\ref{fig:sigmatot}.
One can see that the maximum is about 2.3 mb at $\sqrt{s}\approx
4.1$ GeV. This is close to the result obtained in
\cite{Wong:1999zb,Wong:2001td,Barnes:2003dg}.

\acknowledgements{
We thank David Blaschke and Craig Roberts for useful discussions.
M.A.I. gratefully acknowledges the hospitality and support of
Mainz University during a visit in which some of this work was
conducted. This work was supported in part by a DFG Grant, the
Russian Fund of Basic Research Grant No. 01-02-17200 and the
Heisenberg-Landau Program.}

\newpage

\appendix*
\section{Expressions for form factors}

Here we report the analytical expressions for the form factors
involved in the calculation of the amplitudes $J/\psi+\pi^+\to
D^{\ast\,+} + \overline{D^{\,0}}$ and $J/\psi+\pi^+\to D^{\ast\,+}
+ \overline{D^{\ast\,0}}$, respectively.

\begin{center}
{\boldmath{\underline{$J/\psi+\pi^+\to D^{\ast\,+} +
\overline{D^{\,0}}$}}}
\end{center}

\begin{eqnarray*}
F^{(i)}_{\rm VPPV}(s,t) &=& g_{J/\psi}\,g_D\,g_{D^\ast}\,g_\pi
\int\! d\sigma_\Box \cdot\exp(-w\cdot z_\Box(s,t))\cdot N_{\rm
VPPV}^{(i)}
\\
&&\\
N_{\rm VPPV}^{(1)} &=&
       - m_q\,m_c\,w^2
       + m_q^2\,w^2 \, ( 2\,d_1\,d_3 - 2\,d_2\,d_3 - d_3 )
       + m_c^2\,w^2 \, (  - 1 + d_1 - d_2 )
\\
&&
       + m^2_{J/\psi}\,w^2 \, (  - 4\,d_1\,d_2\,d_3 - 2\,d_1\,d_2\,d_3^2
- 2\,d_1\,d_2 - 2\,d_1\,d_2^2\,d_3
\\
&&
  - d_1\,d_2^2 - 2\,d_1\,d_3 - d_1\,d_3^2 - d_1
\\
&& + 4\,d_1^2\,d_2\,d_3 + 2\,d_1^2\,d_2 + 4\,d_1^2\,d_3
\\
&& + 2\,d_1^2\,d_3^2 + 2\,d_1^2 - 2\,d_1^3\,d_3 - d_1^3 )
\\
&&
       + m^2_D\,w^2 \, (  - 6\,d_1\,d_2\,d_3 + 2\,d_1\,d_2\,d_3^2
- 2\,d_1\,d_2 - 4\,d_1\,d_2^2\,d_3 - 2\,d_1\,d_2^2 - 2\,d_1\,d_3
\\
&& + 2\,d_1\,d_3^2 + 2\,d_1^2\,d_2\,d_3 + d_1^2\,d_2 +
2\,d_1^2\,d_3
\\
&& + 2\,d_2\,d_3 - 3\,d_2\,d_3^2 + d_2 + 4\,d_2^2\,d_3 -
2\,d_2^2\,d_3^2 + 2\,d_2^2 + 2\,d_2^3\,d_3 + d_2^3 - d_3^2 )
\\
&&
       + m^2_D\,w^2 \, (  - 2\,d_1\,d_3^3 + d_1^2\,d_3 + 2\,d_1^2\,d_3^2
- 2\,d_2\,d_3
\\
&& - 2\,d_2\,d_3^2 + 2\,d_2\,d_3^3 - d_2^2\,d_3 - 2\,d_2^2\,d_3^2
- d_3 - d_3^2+ d_3^3 )
\\
&&
 + u\,w^2 \, ( 3\,d_1\,d_2\,d_3 + 2\,d_1\,d_2 + 2\,d_1\,d_2^2\,d_3
+ d_1\,d_2^2 + d_1\,d_3 -
         2\,d_1^2\,d_2\,d_3 - d_1^2\,d_2 - 2\,d_1^2\,d_3 )
\\
&&
 + s\,w^2 \, ( d_1\,d_2\,d_3 + 2\,d_1\,d_2\,d_3^2 + d_1\,d_3
+ d_1\,d_3^2 - d_1^2\,d_3 - 2\,
         d_1^2\,d_3^2 )
\\
&&
  + t\,w^2 \, (  - d_1\,d_2\,d_3 - 2\,d_1\,d_2\,d_3^2 - d_1\,d_3
- 2\,d_1\,d_3^2 + 2\,d_2\,d_3
\\
&&
          + 3\,d_2\,d_3^2 + d_2^2\,d_3 + 2\,d_2^2\,d_3^2 + d_3 + d_3^2 )
\\
&&
       + w \, (  - 3 - 8\,v\,d_1\,d_3 - 3\,v\,d_1 + 8\,v\,d_2\,d_3
+ 3\,v\,d_2 + 5\,v\,d_3
\\
&& + 3\,v + 8\,d_1\,d_3 + 3\,d_1 - 8\,d_2\,d_3 - 3\,d_2 - 5\,d_3 ),
\\
&&\\
N_{\rm VPPV}^{(2)} &=&
       + m_q^2\,w^2 \, (  - 2\,d_1\,d_2 + d_2 + 2\,d_2^2 )
\\
&&
       + m^2_{J/\psi}\,w^2 \, ( d_1\,d_2\,d_3 + 2\,d_1\,d_2^2\,d_3
+ 3\,d_1\,d_2^2 + 2\,d_1\,d_2^3 - 2\,d_1^2\,
\\
&& d_2\,d_3 - 3\,d_1^2\,d_2 - 4\,d_1^2\,d_2^2 + 2\,d_1^3\,d_2 )
\\
&&
       + m^2_D\,w^2 \, (  - 2\,d_1\,d_2\,d_3 + 2\,d_1\,d_2
- 2\,d_1\,d_2^2\,d_3 + 7\,d_1\,d_2^2
\\
&& + 4\,d_1\,d_2^3 - 2\,d_1^2\,d_2 - 2\,d_1^2\,d_2^2 + d_2\,d_3 -
d_2 + 3\,d_2^2\,d_3 - 4\,d_2^2 + 2\,d_2^3\,d_3 - 5\,d_2^3 -
2\,d_2^4 )
\\
&& + m^2_\pi\,w^2 \, (  - d_1\,d_2\,d_3 + 2\,d_1\,d_2\,d_3^2 -
2\,d_1\,d_2 - 2\,d_1^2\,d_2\,d_3
\\
&& + 3\,d_2\,d_3 - d_2\,d_3^2 + 3\,d_2^2\,d_3 - 2\,d_2^2\,d_3^2 +
2\,d_2^3\,d_3 + d_3 )
\\
&&
       + u\,w^2 \, (  - d_1\,d_2 - 3\,d_1\,d_2^2 - 2\,d_1\,d_2^3
+ 2\,d_1^2\,d_2 + 2\,d_1^2\,d_2^2 )
\\
&&
       + s\,w^2 \, (  - d_1\,d_2\,d_3 + d_1\,d_2 - 2\,d_1\,d_2^2\,d_3
+ 2\,d_1^2\,d_2\,d_3 )
\\
&&
       + t\,w^2 \, ( 2\,d_1\,d_2\,d_3 + d_1\,d_2 + 2\,d_1\,d_2^2\,d_3
\\
&& - d_2\,d_3 - 3\,d_2^2\,d_3 -
         2\,d_2^3\,d_3 )
\\
&&
       + w \, ( 1 + 8\,v\,d_1\,d_2 + 2\,v\,d_1 - 7\,v\,d_2 - 8\,v\,d_2^2
- v - 8\,d_1\,d_2 - 2\,d_1 + 7\,d_2 + 8\,d_2^2 ),
\\
&&\\
N_{\rm VPPV}^{(3)} &=&
       + m_q\,m_c\,w^2 \, ( 2\,d_3 )
       + m_q^2\,w^2 \, (  - 2\,d_2\,d_3 + 2\,d_3^2 )
       + m_c^2\,w^2 \, (  - 1 - d_2 )
\\
&&
       + m^2_{J/\psi}\,w^2 \, (  - d_1\,d_2\,d_3 - d_1\,d_2
- 2\,d_1\,d_2^2\,d_3 - d_1\,d_2^2 + d_1\,d_3
\\
&& + 4\,d_1\,d_3^2 + 2\,d_1\,d_3^3 - d_1 + 2\,d_1^2\,d_2\,d_3 +
d_1^2\,d_2 - 2\,d_1^2\,d_3 - 2\,d_1^2\,d_3^2 + d_1^2 )
\\
&&
       + m^2_D\,w^2 \, ( 2\,d_1\,d_2\,d_3^2 - d_1\,d_2 - 2\,d_1\,d_2^2\,d_3
\\
&& - d_1\,d_2^2 + d_1\,d_3 + 2\,d_1\,d_3^2 - d_2\,d_3 -
4\,d_2\,d_3^2
\\
&& + 2\,d_2\,d_3^3 + d_2^2\,d_3 - 4\,d_2^2\,d_3^2 +
         d_2^2 + 2\,d_2^3\,d_3 + d_2^3 + 2\,d_3^3 )
\\
&&
       + m^2_\pi\,w^2 \, (  - d_1\,d_2\,d_3 - 2\,d_1\,d_2\,d_3^2
+ 2\,d_1\,d_3^2 + 2\,d_1\,d_3^3 - d_2\,d_3
\\
&& + d_2\,d_3^2 + 4\,d_2\,d_3^3 - d_2^2\,d_3 - 2\,d_2^2\,d_3^2 +
d_3^2 - 2\,d_3^4 )
\\
&&
       + u\,w^2 \, (  - 2\,d_1\,d_2\,d_3^2 + d_1\,d_2
+ 2\,d_1\,d_2^2\,d_3 + d_1\,d_2^2 - d_1\,d_3
          - 2\,d_1\,d_3^2 )
\\
&&
       + s\,w^2 \, ( d_1\,d_2\,d_3 + 2\,d_1\,d_2\,d_3^2 - 2\,d_1\,d_3^2
- 2\,d_1\,d_3^3 )
\\
&&
       + t\,w^2 \, ( d_2\,d_3 - 2\,d_2\,d_3^3 + d_2^2\,d_3 + 2\,d_2^2\,d_3^2
- 2\,d_3^2 - 2\,d_3^3 )
\\
&&
       + w \, (  - 1 + 8\,v\,d_2\,d_3 + 3\,v\,d_2 - 4\,v\,d_3 - 8\,v\,d_3^2
+ v - 8\,d_2\,d_3 - 3\,d_2 + 4\,d_3 + 8\,d_3^2 ),
\\
&&\\
N_{\rm VPPV}^{(4)} &=&
       + m_q\,m_c\,w^2 \, (  - 2\,d_2 )
       + m_q^2\,w^2 \, (  - 2\,d_2\,d_3 + 2\,d_2^2 )
       + m_c^2\,w^2 \, ( 1 + d_2 )
\\
&&
       + m^2_{J/\psi}\,w^2 \, (  - 3\,d_1\,d_2\,d_3 - 2\,d_1\,d_2\,d_3^2
- d_1\,d_2 + d_1\,d_2^2
\\
&& + 2\,d_1\,d_2^3 + d_1\,d_3 + d_1 + 2\,d_1^2\,d_2\,d_3 +
d_1^2\,d_2 - 2\,d_1^2\,d_2^2 - d_1^2 )
\\
&&
       + m^2_D\,w^2 \, (  - 2\,d_1\,d_2\,d_3 - 2\,d_1\,d_2^2\,d_3
+ d_1\,d_2^2 + 2\,d_1\,d_2^3 + d_2\,d_3 - 2\,d_2\,d_3^2 +
5\,d_2^2\,d_3
\\
&& - 2\,d_2^2\,d_3^2 - d_2^2 + 4\,d_2^3\,d_3 - 3\,d_2^3
          - 2\,d_2^4 )
\\
&&
       + m^2_\pi\,w^2 \, (  - d_1\,d_2\,d_3 - 2\,d_1\,d_2\,d_3^2 - d_1\,d_2
+ 2\,d_1\,d_2^2\,d_3 + d_1\,d_3
\\
&&
          + d_2\,d_3 - d_2\,d_3^2 + 2\,d_2\,d_3^3 + d_2^2\,d_3
- 4\,d_2^2\,d_3^2 + 2\,d_2^3\,d_3 - d_3^2 )
\\
&&
       + u\,w^2 \, ( 2\,d_1\,d_2\,d_3 + 2\,d_1\,d_2^2\,d_3 - d_1\,d_2^2
- 2\,d_1\,d_2^3 )
\\
&&
       + s\,w^2 \, ( d_1\,d_2\,d_3 + 2\,d_1\,d_2\,d_3^2 + d_1\,d_2
- 2\,d_1\,d_2^2\,d_3 - d_1\,d_3 )
\\
&&
       + t\,w^2 \, ( d_2\,d_3 + 2\,d_2\,d_3^2 - d_2^2\,d_3 + 2\,d_2^2\,d_3^2
- 2\,d_2^3\,d_3 )
\\
&&
       + w \, ( 1 + 8\,v\,d_2\,d_3 - v\,d_2 - 8\,v\,d_2^2 + 2\,v\,d_3 - v
- 8\,d_2\,d_3 + d_2 + 8\,d_2^2 - 2\,d_3 ),
\\
&&\\
N_{\rm VPPV}^{(5)} &=&
       + m_q\,m_c\,m^2_{J/\psi}\,w^2 \, (  - 1/2 )
       + m_q\,m_c\,m^2_\pi\,w^2 \, (  - 1/2 )
\\
&&
       + m_q\,m_c\,s\,w^2 \, ( 1/2 )
       + m_q^2\,m_c^2\,w^2 \, ( 1 )
\\
&&
       + m_q^2\,m^2_{J/\psi}\,w^2 \, ( d_1\,d_2 + d_1\,d_3 + d_1 - d_1^2
- 1/2\,d_2 - 1/2\,d_3 )
\\
&&
       + m_q^2\,m^2_D\,w^2 \, ( d_1\,d_2 + d_2\,d_3 - 1/2\,d_2 - d_2^2 )
\\
&&
       + m_q^2\,m^2_\pi\,w^2 \, ( d_1\,d_3 + d_2\,d_3 - 1/2\,d_3 - d_3^2 )
\\
&&
       + m_q^2\,u\,w^2 \, (  - d_1\,d_2 + 1/2\,d_2 )
\\
&&
       + m_q^2\,s\,w^2 \, (  - d_1\,d_3 + 1/2\,d_3 )
\\
&&
       + m_q^2\,t\,w^2 \, (  - d_2\,d_3 )
\\
&&
       + m_q^2\,w \, ( 1 - v )
\\
&&
       + m_c^2\,m^2_{J/\psi}\,w^2 \, ( d_1\,d_2 + d_1\,d_3 + 3/2\,d_1
- d_1^2 )
\\
&&
       + m_c^2\,m^2_D\,w^2 \, (  - 1/2 + d_1\,d_2 + d_1 + d_2\,d_3 - 3/2\,d_2 - d_2^2 + d_3 )
\\
&&
       + m_c^2\,m^2_\pi\,w^2 \, ( 1/2 + d_1\,d_3 + 1/2\,d_1 + d_2\,d_3
+ 1/2\,d_2 - d_3^2 )
\\
&&
       + m_c^2\,u\,w^2 \, (  - d_1\,d_2 - d_1 )
\\
&&
       + m_c^2\,s\,w^2 \, (  - d_1\,d_3 - 1/2\,d_1 )
\\
&&
       + m_c^2\,t\,w^2 \, (  - 1/2 - d_2\,d_3 - 1/2\,d_2 - d_3 )
\\
&&
       + m_c^2\,w \, ( 2 - 2\,v )
\\
&&
       + m^2_{J/\psi}\,m^2_D\,w^2 \, ( d_1\,d_2\,d_3 + 2\,d_1\,d_2\,d_3^2
- 4\,d_1\,d_2 - 5\,d_1\,d_2^2
\\
&& - 2\,d_1\,d_2^3 + d_1\,d_3^2 - 1/2\,d_1 + 11/2\,d_1^2\,d_2 +
4\,d_1^2\,d_2^2 + 3/2\,d_1^2
\\
&& - 2\,d_1^3\,d_2 - d_1^3 - 1/2\,d_2\,d_3^2 + d_2 + 3/2\,d_2^2 +
1/2\,d_2^3 - 1/2\,d_3^2 )
\\
&&
       + m^2_{J/\psi}\,m^2_\pi\,w^2 \, ( 2\,d_1\,d_2\,d_3 + 2\,d_1\,d_2^2\,d_3
+ 1/2\,d_1\,d_2^2 - 7/2\,d_1\,d_3^2
\\
&&
          - 2\,d_1\,d_3^3 + 1/2\,d_1 + 7/2\,d_1^2\,d_3 + 4\,d_1^2\,d_3^2
- 2\,d_1^3\,d_3 - 1/2\,
         d_1^3 - d_2\,d_3 - 1/2\,d_2^2\,d_3 - d_3 + 1/2\,d_3^3 )
\\
&&
       + m^2_{J/\psi}\,u\,w^2 \, ( d_1\,d_2\,d_3 + 3/2\,d_1\,d_2 + d_1\,d_2^2 + 1/2\,d_1\,d_3 - 2\,d_1^2\,d_2\,d_3
\\
&& - 4\,d_1^2\,d_2 - 2\,d_1^2\,d_2^2 - d_1^2\,d_3 - d_1^2 +
2\,d_1^3\,d_2 + d_1^3 )
\\
&&
       + m^2_{J/\psi}\,s\,w^2 \, ( d_1\,d_2\,d_3 + 1/2\,d_1\,d_2
+ 3/2\,d_1\,d_3 + d_1\,d_3^2
\\
&& - 2\,d_1^2\,d_2\,d_3 - 1/2\,d_1^2\,d_2 - 7/2\,d_1^2\,d_3 -
2\,d_1^2\,d_3^2 - 1/2\,d_1^2 + 2\,d_1^3\,d_3 + 1/2\,d_1^3 )
\\
&&
       + m^2_{J/\psi}\,t\,w^2 \, (  - 4\,d_1\,d_2\,d_3 - 2\,d_1\,d_2\,d_3^2
- 1/2\,d_1\,d_2 - 2\,d_1\,d_2^2\,d_3
\\
&& - 1/2\,d_1\,d_2^2 - 3/2\,d_1\,d_3 - d_1\,d_3^2 - 1/2\,d_1 +
2\,d_1^2\,d_2\,d_3
\\
&& + 1/2\,d_1^2\,d_2 + d_1^2\,d_3 + 1/2\,d_1^2 + d_2\,d_3 +
1/2\,d_2\,d_3^2 + 1/2\,d_2^2\,d_3 + 1/2\,d_3 + 1/2\,d_3^2 )
\\
&&
       + m^2_{J/\psi}\,w \, (  - 3/2 - 5\,v\,d_1\,d_2 - 5\,v\,d_1\,d_3
- 13/2\,v\,d_1 + 5\,v\,d_1^2
\\
&& + 3/2\,v\,d_2 + 3/2\,v\,d_3 + 3/2\,v + 5\,d_1\,d_2 +
5\,d_1\,d_3 + 13/2\,d_1 - 5\,d_1^2 - 3/2\,d_2 - 3/2\,d_3 )
\\
&&
       + m^4_{J/\psi}\,w^2 \, (  - d_1\,d_2\,d_3 - d_1\,d_2
- 1/2\,d_1\,d_2^2 - d_1\,d_3 - 1/2\,d_1\,d_3^2
\\
&&
          + 2\,d_1^2\,d_2\,d_3 + 3\,d_1^2\,d_2 + d_1^2\,d_2^2 + 3\,d_1^2\,d_3
+ d_1^2\,d_3^2
\\
&& + 3/2\,d_1^2 - 2\,d_1^3\,d_2 - 2\,d_1^3\,d_3 - 5/2\,d_1^3 +
d_1^4 )
\\
&&
       + m^2_D\,m^2_\pi\,w^2 \, (  - d_1\,d_2\,d_3 - d_1\,d_3
+ 2\,d_1^2\,d_2\,d_3 + 1/2\,d_1^2\,d_2
\\
&& + d_1^2\,d_3 + 5/2\,d_2\,d_3^2 - 2\,d_2\,d_3^3 + 1/2\,d_2 -
d_2^2\,d_3 + 4\,d_2^2\,d_3^2
\\
&& - 2\,d_2^3\,d_3 - 1/2\,d_2^3 - 1/2\,d_3 - d_3^3 )
\\
&&
       + m^2_D\,u\,w^2 \, (  - 2\,d_1\,d_2\,d_3 + 3/2\,d_1\,d_2
- 2\,d_1\,d_2^2\,d_3
\\
&& + 7/2\,d_1\,d_2^2 + 2\,d_1\,d_2^3 - 2\,d_1^2\,d_2 -
2\,d_1^2\,d_2^2 + 1/2\,d_2\,d_3 - 1/2\,d_2
\\
&& + 1/2\,d_2^2\,d_3 - d_2^2 - 1/2\,d_2^3 )
\\
&&
       + m^2_D\,s\,w^2 \, ( 2\,d_1\,d_2\,d_3 - 2\,d_1\,d_2\,d_3^2
+ 1/2\,d_1\,d_2 + 2\,d_1\,d_2^2\,d_3
\\
&&
          + 1/2\,d_1\,d_2^2 + d_1\,d_3 - d_1\,d_3^2 - 2\,d_1^2\,d_2\,d_3
- 1/2\,d_1^2\,d_2 - d_1^2\,d_3
\\
&& - 1/2\,d_2\,d_3 + 1/2\,d_2\,d_3^2 - 1/2\,d_2 - 1/2\,d_2^2\,d_3
- 1/2\,d_2^2 + 1/2\,d_3^2 )
\\
&&
       + m^2_D\,t\,w^2 \, (  - 2\,d_1\,d_2\,d_3 - 1/2\,d_1\,d_2
- 2\,d_1\,d_2^2\,d_3
\\
&& - 1/2\,d_1\,d_2^2 + 1/2\,d_2\,d_3 - 2\,d_2\,d_3^2 +
5/2\,d_2^2\,d_3 - 2\,d_2^2\,d_3^2
\\
&& + 1/2\,d_2^2+ 2\,d_2^3\,d_3 + 1/2\,d_2^3 )
\\
&&
       + m^2_D\,w \, (  - 1 - 5\,v\,d_1\,d_2 - 2\,v\,d_1 - 5\,v\,d_2\,d_3 + 4\,v\,d_2
\\
&& + 5\,v\,d_2^2- 2\,v\,d_3 + v + 5\,d_1\,d_2 + 2\,d_1 +
5\,d_2\,d_3 - 4\,d_2 - 5\,d_2^2 + 2\,d_3 )
\\
&&
       + m^4_D\,w^2 \, ( 2\,d_1\,d_2\,d_3 - d_1\,d_2 + 2\,d_1\,d_2^2\,d_3 - 3\,d_1\,d_2^2
- 2\,d_1\,d_2^3 + d_1^2\,d_2 + d_1^2\,d_2^2
\\
&& - d_2\,d_3 + d_2\,d_3^2 + 1/2\,d_2 - 3\,d_2^2\,d_3 +
         d_2^2\,d_3^2 + 3/2\,d_2^2 - 2\,d_2^3\,d_3 + 2\,d_2^3 + d_2^4 )
\\
&&
       + m^2_\pi\,u\,w^2 \, ( 2\,d_1\,d_2\,d_3^2 - 2\,d_1\,d_2^2\,d_3 - 1/2\,d_1\,d_2^2
\\
&& + 1/2\,d_1\,d_3 + d_1\,d_3^2 - 2\,d_1^2\,d_2\,d_3 -
1/2\,d_1^2\,d_2 - d_1^2\,d_3 + 1/2\,d_2\,d_3
\\
&& - 1/2\,d_2\,d_3^2 + 1/2\,d_2^2\,d_3 + 1/2\,d_3 )
\\
&&
       + m^2_\pi\,s\,w^2 \, (  - d_1\,d_2\,d_3 - 2\,d_1\,d_2\,d_3^2 + 1/2\,d_1\,d_3
\\
&& + 3/2\,d_1\,d_3^2 + 2\,d_1\,d_3^3 - d_1^2\,d_3 -
2\,d_1^2\,d_3^2 + 1/2\,d_2\,d_3 + 1/2\,d_2\,d_3^2 + 1/2\,d_3 -
1/2\,d_3^3 )
\\
&&
       + m^2_\pi\,t\,w^2 \, (  - d_1\,d_2\,d_3 - 2\,d_1\,d_2\,d_3^2 - 1/2\,d_1\,d_3 - d_1\,d_3^2
\\
&& - 1/2\,d_2\,d_3 + 2\,d_2\,d_3^3 - d_2^2\,d_3 - 2\,d_2^2\,d_3^2
+ 1/2\,d_3 + d_3^2 + d_3^3 )
\\
&&
       + m^2_\pi\,w \, (  - 1/2 - 5\,v\,d_1\,d_3 - 3/2\,v\,d_1
- 5\,v\,d_2\,d_3 - 3/2\,v\,d_2
\\
&& + 5/2\,v\,d_3 + 5\,v\,d_3^2 + 1/2\,v + 5\,d_1\,d_3
\\
&& + 3/2\,d_1 + 5\,d_2\,d_3 + 3/2\,d_2 - 5/2\,d_3 - 5\,d_3^2 )
\\
&&
       + m^4_\pi\,w^2 \, ( d_1\,d_2\,d_3 + 2\,d_1\,d_2\,d_3^2 - d_1\,d_3^2 - 2\,d_1\,d_3^3
+ 1/2\,d_1^2\,d_3
\\
&& + d_1^2\,d_3^2 - d_2\,d_3^2 - 2\,d_2\,d_3^3 + 1/2\,d_2^2\,d_3 +
d_2^2\,d_3^2
          - 1/2\,d_3 - 1/2\,d_3^2 + 1/2\,d_3^3 + d_3^4 )
\\
&&
       + u\,s\,w^2 \, (  - d_1\,d_2\,d_3 - 1/2\,d_1\,d_2 - 1/2\,d_1\,d_3 + 2\,d_1^2\,d_2\,d_3
          + 1/2\,d_1^2\,d_2 + d_1^2\,d_3 )
\\
&&
       + u\,t\,w^2 \, ( 2\,d_1\,d_2\,d_3 + 1/2\,d_1\,d_2 + 2\,d_1\,d_2^2\,d_3 + 1/2\,d_1\,d_2^2
          - 1/2\,d_2\,d_3 - 1/2\,d_2^2\,d_3 )
\\
&&
       + u\,w \, ( 1/2 + 5\,v\,d_1\,d_2 + 2\,v\,d_1 - 3/2\,v\,d_2 - 1/2\,v - 5\,d_1\,d_2 - 2\,
         d_1 + 3/2\,d_2 )
\\
&&
       + u^2\,w^2 \, (  - 1/2\,d_1\,d_2 - 1/2\,d_1\,d_2^2 + d_1^2\,d_2 + d_1^2\,d_2^2 )
\\
&&
       + s\,t\,w^2 \, ( d_1\,d_2\,d_3 + 2\,d_1\,d_2\,d_3^2 + 1/2\,d_1\,d_3 + d_1\,d_3^2 - 1/2
         \,d_2\,d_3 - 1/2\,d_2\,d_3^2 - 1/2\,d_3 - 1/2\,d_3^2 )
\\
&&
       + s\,w \, ( 1 + 5\,v\,d_1\,d_3 + 3/2\,v\,d_1 - 3/2\,v\,d_3 - v - 5\,d_1\,d_3 - 3/2\,d_1
          + 3/2\,d_3 )
\\
&&
       + s^2\,w^2 \, (  - 1/2\,d_1\,d_3 - 1/2\,d_1\,d_3^2 + 1/2\,d_1^2\,d_3 + d_1^2\,d_3^2 )
\\
&&
       + t\,w \, (  - 1/2 + 5\,v\,d_2\,d_3 + 3/2\,v\,d_2 + 2\,v\,d_3 + 1/2\,v - 5\,d_2\,d_3
          - 3/2\,d_2 - 2\,d_3 )
\\
&&
       + t^2\,w^2 \, ( 1/2\,d_2\,d_3 + d_2\,d_3^2 + 1/2\,d_2^2\,d_3 + d_2^2\,d_3^2 )
\\
&&
       + 3 - 6\,v + 3\,v^2;
\end{eqnarray*}

\begin{eqnarray*}
F^{(1a)}_{\rm PVP}(t) &=& g_\pi\,g_{D^\ast}\,g_D\,\int\!
d\sigma_\triangle \cdot \exp(-w z_\triangle(t))\,
\cdot[w\,m_q\,m_c \, (  - 1 )
       + w\,m_q^2 \, (  - b_1 )
\\
&&
       + w\,m^2_\pi \, ( b_1^2\,b_2 - b_1^2 + b_1^3 )
       + w\,t \, ( b_1\,b_2^2 + b_1^2\,b_2 - b_2 + b_2^2 )
\\
&&
       + w\,m^2_D \, (  - b_1^2\,b_2 )
       + v \, ( 1 + 3\,b_1 )
       - 1 - 3\,b_1],
\\
F^{(2a)}_{\rm PVP}(t) &=& g_{J/\psi}\,g_{D^\ast}\,g_D\,\int\!
d\sigma_\triangle \cdot \exp(-w z_\triangle(t))\, \cdot[w\,m_q^2
\, ( 1 - b_2 )
\\
&&
    + w\,m^2_\pi \, (  - 3\,b_1\,b_2 + b_1\,b_2^2 + b_1 + b_1^2\,b_2
- b_1^2 + b_2 - b_2^2 )
\\
&&
       + w\,t \, (  - b_1\,b_2 + b_1\,b_2^2 + b_2 - 2\,b_2^2 + b_2^3 )
\\
&&
       + w\,m^2_D \, ( b_1\,b_2 - b_1\,b_2^2 - b_2 + b_2^2 )
\\
&&
       + v \, (  - 2 + 3\,b_2 )
       + 2 - 3\,b_2],
\\
&&\\
F^{(b)}_{\rm PVV}(u) &=& g_\pi\,g^2_{D^\ast}\int\!
d\sigma_\triangle \cdot \exp(-w z_\triangle(u))\, \cdot[w\,
(m_q\,(b_2 - 1) - m_c\,b_2)],
\\
&&\\
F^{(b)}_{\rm VPV}(u) &=& g_{J/\psi}\,g_D\,g_{D^\ast}\,\int\!
d\sigma_\triangle \cdot \exp(-w z_\triangle(u)) \cdot [w\,( m_c\,(
b_2 - 1) - m_q\, b_2 )],
\\
&&\\
&&\\
F^{(c)}_{\rm PVP}(u) &=& g_\pi\,g_{D^\ast}\,g_D\,\int\!
d\sigma_\triangle \cdot \exp(-w z_\triangle(u))\, \cdot[w\,(
(-m_c\,m_q - m_q^2\,b_1)
\\
&&
       + m^2_\pi \, b_1^2\,(b_2 - 1 + b_1 )
\\
&&
       + m^2_D\,b_2\, ( b_1\,b_2 + b_1^2 - 1 + b_2 )
\\
&&
       - u\,b_1^2\,b_2)  -(1-v)\, (  1 + 3\,b_1 )],
\\
F^{(c)}_{\rm VPP}(u) &=& g_{J/\psi}\,g_D^2\,\int\!
d\sigma_\triangle \cdot \exp(-w z_\triangle(u))\,
\cdot[w\,(-2\,m_c\,m_q \, b_2
       +m_c^2 \, (  - 1 + b_2 )
\\
&&
       + m^2_{J/\psi} \, b_1\,(b_2^2 - 1 + b_1\,b_2 + b_1 )
\\
&&
       +m^2_D\, b_2^2\, ( b_1 - 1 + b_2 ) - u\,b_1\,b_2^2)
\\
&&
       -(1-v)\, (1 + 3\,b_2)]
\end{eqnarray*}

\vspace{1cm}

\begin{center}
{\boldmath{\underline{$J/\psi+\pi^+\to D^{\ast\,+} +
\overline{D^{\ast\,0}}$}}}
\end{center}

\vspace{-0.5cm}
\begin{eqnarray*}
F^{(i)}_{\rm VVPV}(s,t) &=& g_{J/\psi}\,g^2_{D^\ast}\,g_\pi \int\!
d\sigma_\Box \cdot\exp(-w\cdot z_\Box(s,t))\cdot N_{\rm
VVPV}^{(i)}
\\
&&\\
N_{\rm VVPV}^{(1)} &=& -\,w^2  m_q\, d_2, \\
N_{\rm VVPV}^{(2)} &=& w^2  m_c\, d_1, \\
N_{\rm VVPV}^{(3)} &=& w^2 m_c\, (1-d_1+d_2), \\
N_{\rm VVPV}^{(4)} &=& w^2 m_c\, (1-2\, d_1+d_2), \\
N_{\rm VVPV}^{(5)} &=& - w^2 m_c\, d_1, \\
N_{\rm VVPV}^{(6)} &=& w^2 \left( m_c\, (1-d_1+d_2) - m_q\, d_3\right), \\
N_{\rm VVPV}^{(7)} &=& - w^2 m_q\, d_3, \\
N_{\rm VVPV}^{(8)} &=&
w^2 d_3\,\left(m_q \, (2\, d_1-1) - 2\, m_c\, d_1\right), \\
N_{\rm VVPV}^{(9)} &=&
w^2 d_2\,\left( m_q \,(1-2\, d_1) + 2\, m_c\, d_1\right), \\
N_{\rm VVPV}^{(10)} &=& w^2 ( m_q\, d_2 - m_c\, d_1), \\
N_{\rm VVPV}^{(11)} &=&
w^2 \left(m_q\, d_2  + m_c\, (d_1- d_2-1)\right), \\
N_{\rm VVPV}^{(12)} &=& - w^2 m_c\, d_1, \\
N_{\rm VVPV}^{(13)} &=& w^2 m_c\, (d_1-d_2 -1), \\
N_{\rm VVPV}^{(14)}  &=&
2\,w^2 d_3\, \left(m_q\, d_2  - m_c\, (1+d_2)\right), \\
N_{\rm VVPV}^{(15)} &=&
-\,2\, w^2 d_2\, \left( m_q\, d_2  - m_c\, (1+d_2)\right), \\
N_{\rm VVPV}^{(16)} &=& w^2 m_c\, \left\{ -m_q^2+m_{J/\psi}^2\,
d_1\,\left(d_1-\frac{3}{2}-d_2-d_3\right) \right.
\\
&+& m_{D^\ast}^2\,\left(\frac{1}{2} -d_1d_2-d_1-d_2d_3
         +\frac{3}{2}\, d_2 +d_2^2 - d_3 \right)
\\
&+& m_\pi^2\,\left(-\frac{1}{2} -d_1d_3-\frac{1}{2}\, d_1-d_2d_3
               -\frac{1}{2}\, d_2 +d_3^2 \right)
\\
&+& \left. u\, d_1\,(1+d_2)+(p_1+s\,
d_1\,\left(\frac{1}{2}+d_3\right)
   +t\,\left(\frac{1}{2}\,(1+d_2)+d_3\,(1+d_2)\right)
\right\}
\\
&-& 2\, m_c\, w(1-v), \\
N_{\rm VVPV}^{(17)} &=& w^2 \,m_q\, \left\{ -\,m_c^2  +
m_{J/\psi}^2\,\left(  - d_1 d_2 - d_1d_3 - d_1 + d_1^2 +
                   \frac{1}{2}\,(d_2 + d_3) \right)
\right.
\\
&+& m_{D^\ast}^2\, d_2\, \left(d_2-d_1-d_3+\frac{1}{2}\right)
+m_\pi^2\, d_3\,\left(d_3-d_1-d_2+\frac{1}{2}\right)
\\
&+& \left. u\, d_2 \left(d_1-\frac{1}{2}\right)
+s\,d_3\,\left(d_1-\frac{1}{2}\right)
          + t\, d_2\,d_3
\right\}
\\
&-& w\, (1-v)\, (m_q+m_c).
\end{eqnarray*}

\begin{eqnarray*}
F^{(a)}_{\rm PVV}(t) &=& g_\pi\,g^2_{D^\ast}\,\int\!
d\sigma_\triangle \cdot \exp(-w z_\triangle(t))\, \cdot [w\,
(m_q\, (b_2 - 1) -  m_c\, b_2)],
\\
&&\\
F^{(1a)}_{\rm VVV}(t) &=& g_{J/\psi}\,g^2_{D^\ast}\,\int\!
d\sigma_\triangle \cdot \exp(-w z_\triangle(t))\,
\cdot[4\,w\,b_1\,b_2\,(b_1+b_2 - 1)],
\\
F^{(2a)}_{\rm VVV}(t) &=& g_{J/\psi}\,g^2_{D^\ast}\,\int
d\sigma_\triangle \cdot \exp(-w z_\triangle(t))\, \cdot
[2\,w\,b_2\,(b_1 - 2\,b_1\, b_2 - 1 + 3\, b_2 - 2\, b_2^2)],
\\
F^{(3a)}_{\rm VVV}(t) &=& g_{J/\psi}\,g^2_{D^\ast}\,\int\!
d\sigma_\triangle \cdot \exp(-w z_\triangle(t)) \cdot [  w\, ( -
m_c\,(m_q + m_c b_1)
   + m^2_{J/\psi}\,b_1^2\,(b_1+b_2-1)
\\
&+& t\,b_2\,(b_1^2 + b_1 b_2 +b_2-1)
   - m^2_D\,  b_1^2\, b_2) -(1-v)(b_1+1)],
\\
F^{(4a)}_{\rm VVV}(t) &=& g_{J/\psi}\,g^2_{D^\ast}\,\int\!
d\sigma_\triangle \cdot \exp(-w z_\triangle(t)) \cdot [ w\,(
 m_c^2 (d2 -1)
+m^2_{J/\psi}\,(3\, b_1\,b_2 - b_1\, b_2^2 -b_1
    -  b_1^2\,b_2 +  b_1^2 -  b_2 +  b_2^2)
\\
&+& t\,( b_1\,b_2   -  b_1\,b_2^2 -  b_2 + 2\, b_2^2 -  b_2^3)
   - m^2_D\, (b_1\,b_2  +  b_1\,b_2^2 +  b_2 -  b_2^2))
   + (1-v)\,(b_2 - 1)],
\\
F^{(5a)}_{\rm VVV}(t) &=& g_{J/\psi}\,g^2_{D^\ast}\,\int\!
d\sigma_\triangle \cdot \exp(-w z_\triangle(t)) \cdot
[w\,(m_c\,(m_q - m_c b_1 - m_c b_2 + m_c)
\\
&+&m^2_{J/\psi}\,b_1\, (- 2\, b_2 +  b_2^2 +1 + 2\,b_1\,b_2 -
2\,b_1 + b_1^2)
\\
&+& t\,b_2\,(- 2\, b_1 + 2\, b_1\,b_2 +  b_1^2 + 1
        - 2\, b_2 +  b_2^2)
+m^2_D\,b_1\,b_2\,( 2 - b_2 -  b_1)) +(1-v)\,(2-b_1 -b_2)],
\\
F^{(6a)}_{\rm VVV}(t) &=& g_{J/\psi}\,g^2_{D^\ast}\int\!
d\sigma_\triangle \cdot \exp(-w z_\triangle(t)) \cdot [w\,(
 m_c\,(2\, m_q\,b_2 - m_c\, b_2 + m_c)
   +m^2_{J/\psi}\,b_1\,(-b_2^2 + 1 -  b_1\, b_2  -  b_1)
\\
&+& t\,b_2\,(-  b_1\, b_2 +  b_2 -  b_2^2)
   + m^2_D\, b_1\, b_2^2)
+ (1-v)\,(b_2 + 1)],
\\
&&\\
F^{(c)}_{\rm PPV}(t) &=& g_\pi\,g_D\,g_{D^\ast}\int\!
d\sigma_\triangle \cdot \exp(-w z_\triangle(t))\, \cdot[ w\,(m_q\,
(m_c - m_q b_1 - m_q b_2 + m_q)
\\
&+& m^2_{J/\psi}\,b_1\,(1-2 b_1-2 b_2 + b_2^2 +2 b_1 b_2+b_1^2)
\\
&+& t\,b_2\,(1- 2 b_1 - 2 b_2 +b_1^2 + 2 b_1 b_2 +  b_2^2)
 +m^2_D\,b_1\,b_2\,(2 - b_1-b_2))
     +(1-v)\,(4- 3\,b_1 - 3\,b_2 )]
\\
&&\\
F^{(c)}_{\rm VPV}(t) &=& g_{J/\psi}\,g_D\,g_{D^\ast}\,\int\!
d\sigma_\triangle \cdot \exp(-w z_\triangle(t)) \cdot
[w\,(m_c-b_2\,(m_c-m_q))].
\end{eqnarray*}

\end{document}